\documentclass[%
reprint,superscriptaddress,frontmatterverbose,
preprintnumbers,nofootinbib,nobibnotes,
amsmath,amssymb,aps,
prc,
]{revtex4-1}
\usepackage{graphicx} 
\usepackage{dcolumn}
\usepackage{bm}
\usepackage{color}
\usepackage{url}
\def\ket#1{\left|{#1}\right\rangle}

\def\brakket#1#2#3{\left\langle{#1}\middle|{#2}\middle|{#3}\right\rangle}
\def\ve#1{{\bm{#1}}}
\def\nuc#1#2#3{{}^{#2}_{#3}\mathrm{#1}}
\def\urm#1{\scriptstyle{\text{\textrm{\textmd{\textup{#1}}}}}}

\def\avr#1{\left\langle{#1}\right\rangle}

\def\ca#1{{\mathcal{#1}}}
\DeclareMathOperator{\Div}{div}
\let\temp\epsilon
\let\epsilon\varepsilon
\let\varepsilon\temp
\let\temp\relax
\begin{document}
\preprint{RIKEN-QHP-426}
\preprint{RIKEN-iTHEMS-Report-19}
\title{Theoretical study of $ \mathrm{Nb} $ isotope productions by muon capture reaction on $ {}^{100} \mathrm{Mo} $} 
\author{Maureen Ciccarelli}
\affiliation{Nuclear Data Center, Japan Atomic Energy Agency, Tokai, Ibaraki 319-1195, Japan}
\affiliation{Universit\'{e} de Lille, 59000 Lille, France}
\author{Futoshi Minato}
\email{minato.futoshi@jaea.go.jp}
\affiliation{Nuclear Data Center, Japan Atomic Energy Agency, Tokai, Ibaraki 319-1195, Japan}
\author{Tomoya Naito}
\email{naito@cms.phys.s.u-tokyo.ac.jp}
\affiliation{Department of Physics, Graduate School of Science, The University of Tokyo, Tokyo 113-0033, Japan}
\affiliation{RIKEN Nishina Center, Wako 351-0198, Japan}
\date{\today}
\begin{abstract}
  \begin{description}
  \item[Background]
    The isotope $ \nuc{Mo}{99}{} $, the generator of $ \nuc{Tc}{99m}{} $ used for diagnostic imaging, is supplied by extracting from fission fragments of highly enriched uranium in reactors. 
    However, a reactor-free production method of $\nuc{Mo}{99}{}$ is searched over the world from the point of view of nuclear proliferation.
  \item[Purpose]
    Production methods using accelerators have attracted attention. 
    Recently, $ {}^{99} \mathrm{Mo} $ production through a muon capture reaction was proposed and it was found that about $ 50 \, \% $ of $ \nuc{Mo}{100}{} $ turned into $ \nuc{Mo}{99}{} $ through $ {}^{100} \mathrm{Mo} \left( \mu^-, n \right) $ reaction [arXiv:1908.08166].
    However, the detailed physical process of the muon capture reaction is not completely understood.
    We, therefore, study the muon capture reaction of $ \nuc{Mo}{100}{} $ by a theoretical approach.
  \item[Methods]
    We used the proton-neutron quasi-particle random phase approximation to calculate the muon capture rate. 
    The muon wave function is calculated with considering the electronic distribution of the atom and the nuclear charge distribution.
    The particle evaporation process from the daughter nucleus, $ {}^{100} \mathrm{Nb} $, is calculated by the Hauser-Feshbach statistical model.
  \item[Results]
    From the model calculation, about $ 38 \, \% $ of $ \nuc{Mo}{100}{} $ is converted to $ \nuc{Mo}{99}{} $ through the muon capture reaction, which is in a reasonable agreement with the experimental data. 
    It is revealed that negative parity states, especially $ 1^- $ state, 
    play an important role in $ \nuc{Mo}{100}{} \left( \mu^-, n \right) \nuc{Nb}{99}{} $.
    Charged-particle emission is hindered due to its large separation energy and the Coulomb barrier.
    The feasibility of $ {}^{99} \mathrm{Mo} $ production by the muon capture reaction is also discussed. 
  \item[Conclusions]
    Isotope production by the muon capture reaction strongly depends on the nuclear structure. To understand the mechanism, excitation energy functions have to be known microscopically.
    The muon capture reaction has potential to produce $ \nuc{Mo}{99}{} $ if high-flux muon beam is provided.
  \end{description}
\end{abstract} 
\maketitle
%%%%%%%%%%%%%%%%%%%%%%%%%%%%%% 
\section{Introduction}
\label{sec.introduction}
\par
The isotope $ \nuc{Mo}{99}{} $ ($ T_{1/2} = 66 \, \mathrm{h} $) is used as the generator of $ \nuc{Tc}{99m}{} $, 
which is the most widely used radioisotope for medical diagnostic imaging in the world \cite{Noorden2013}.
Currently, most of $ \nuc{Mo}{99}{} $ is produced by fission reactions of highly enriched $ \nuc{U}{235}{} $ (HEU) or low enriched $ \nuc{U}{235}{} $ in nuclear reactors in some countries \cite{Cherry2003}. 
However, the HEU is an issue of public concern in terms of nuclear proliferation, and a special regulation to deal with it obstructs the global expansion of production place. 
Also, some of the reactors producing $ \nuc{Mo}{99}{} $ have been operated for more than 40 years since they were launched. 
A discussion about the decommissioning of those reactors could happen at any time.
In fact, the National Research Universal (NRU) reactor at Chalk River in Canada, which has covered about $ 40 \, \% $ of the world supply before, entered to shut down in 2018.
\par
For those reasons, an alternative reactor-free production method of $ \nuc{Mo}{99}{} $ is searched in order to sustain its stable supply. 
A production method using accelerators is a promising candidate and has attracted attention. 
Several methods through charged-particle reactions, 
such as 
$ \nuc{Mo}{100}{} \left( p, p \, n \right) \nuc{Mo}{99}{} $, 
$ \nuc{Mo}{100}{} \left( d, p \, 2n \right) \nuc{Mo}{99}{} $, and 
$ \nuc{Mo}{100}{} \left(p, 2n \right) \nuc{Tc}{99}{} $,
and photo-disintegration reactions, such as 
$ \nuc{U}{238}{} \left( \gamma, f \right) \nuc{Mo}{99}{} $ and 
$ \nuc{Mo}{100}{} \left( \gamma, n \right) \nuc{Mo}{99}{} $, have been proposed.
In addition to the above reactions, it is also proposed to use high-energy neutrons produced by accelerators \cite{Nagai2009,Nagai2013},
which are suitable to produce $ \nuc{Mo}{99}{} $ through $ \nuc{Mo}{100}{} \left( n, 2n \right) \nuc{Mo}{99}{} $ reaction \cite{Minato2017}. 
\par
It is also possible to produce $ \nuc{Mo}{99}{} $ by using a negative-muon capture reaction (hereafter, we call simply muon capture).
In this approach, $ \nuc{Mo}{99}{} $ is generated through the $ \beta^- $ decay of $ \nuc{Nb}{99}{} $, 
which is produced by $ \nuc{Mo}{100}{} \left( \mu^-, n \right) \nuc{Nb}{99}{} $ reaction.
The muon capture has several advantages in $ \nuc{Mo}{99}{} $ productions as compared to the aforementioned approaches.
First is that we make the best use of a muon resource because muons rapidly lose the kinetic energy in a target material
and form the muonic atom at a high probability captured by one of the orbits of a nucleus \cite{Nagamine2003}. 
Second is that the muon capture deposits a target nucleus high energy of about $ 10 \, \mathrm{MeV} $ on average, which is suitable to emit only a few neutrons, avoiding to produce unnecessary isotopes.
Third is that target samples can be reused efficiently because the muon capture changes the atomic number of nucleus by only one,
and if the daughter nucleus is unstable, it decays back to the original atomic number. 
This point is also important to suppress the impurities of unnecessary isotopes.
\par
Recently, $ \mathrm{Nb} $ isotope mass distributions by 
$ \nuc{Mo}{100}{} \left( \mu^-, xn \right) \nuc{Nb}{100-x}{} $ \cite{Hashim2018} 
and $ \nuc{Mo}{\text{nat}}{} \left( \mu^-, xn \right) \nuc{Nb}{100-x}{} $~\cite{Hashim2019} 
were studied experimentally at MuSIC in the J-PARC Material Life Science Facility (MLF) and MUSE in Osaka Univ., respectively,
where it was shown that about $ 50 \, \% $ of $ \nuc{Mo}{100}{} $ turned into $ \nuc{Nb}{99}{} $ and more than $ 45 \, \% $ into unstable $ \mathrm{Nb} $ isotopes which become $ \mathrm{Mo} $ isotopes eventually by $ \beta^- $-decay. 
In addition, it was observed that charged-particle emissions were strongly hindered.
This fact indicates that the muon capture is a potential candidate for $ \nuc{Mo}{99}{} $ production if a high-flux muon beam would be gained.
\par
In spite of the above experimental measurements, the muon capture reaction is not perfectly understood from the theoretical point of view.
In particular, it is still not clear why the $ \nuc{Mo}{100}{} \left( \mu^-, n \right) \nuc{Nb}{99}{} $ reaction occurs at such high probability.
The $ \mathrm{Nb} $ isotope mass distribution was discussed in Refs.~\cite{Hashim2018, Hashim2019} using a pre-equilibrium and equilibrium (proton) neutron emission model \cite{Ejiri1989}, which gave a good agreement with the experimental data.
However, the model used phenomenological functions for excitation energies of the daughter nucleus, and they could not discuss the details of muon capture reaction in terms of the nuclear structure microscopically. 
\par
To disentangle the physical processes of the muon capture reaction on $ \mathrm{Mo} $ isotopes, a microscopic model is thus needed.
To this end, in this work, we use the proton-neutron quasiparticle random phase approximation ($ pn $-QRPA) \cite{Halbleib1967Nucl.Phys.A98_542}
on the basis of $ \text{Skyrme-Hartree-Fock} + \text{BCS} $ (SHFBCS) \cite{Vautherin1973} for the muon capture.
For the particle evaporation steps, we adopt the Hauser-Feshbach statistical model (HFSM) \cite{HFSM}, which properly considers the energy conservation, the selection-rule based on the nuclear structure, the transmission probabilities of emitted particles, and so on.
The $ pn $-QRPA is able to cover a wide range of nuclei in the nuclear chart and has been used for a systematical calculation of the muon capture in the nuclear chart \cite{Kolbe1994, Kolbe2000, Zinner2006, Marketin2009}.
For our future plan to make a new table of muon capture reactions, the $ pn $-QRPA is thus adopted in our work.
We would like to stress that the nuclear axial deformation is taken into account in our model.
We also consider effects of the electron distribution, as well as the nuclear finite size, to the muon wave function and its eigenenergy, i.e., binding energy.
\par
In addition to the theoretical study on the muon capture reaction, we also assess the feasibility of $ \nuc{Mo}{99}{} $ production by the muon capture reaction in some muon beam facilities available in the world. 
The production efficiency of $ \nuc{Mo}{99}{} $ by the muon capture reaction is discussed in terms of electric energy, compared with other production approaches.
\par
This paper is organized as follows. 
In Sec.~\ref{sec.Theory}, the theoretical framework used in this work is described. 
In Sec.~\ref{sec.result}, results of the muon capture reaction on $ \nuc{Mo}{100}{} $ and other $ \mathrm{Mo} $ isotopes are given, and the detail about the nuclear structure effects on the muon capture reaction is discussed.
In Sec.~\ref{sec.economical}, we assess the feasibility of $ \nuc{Mo}{99}{} $ production by the muon capture reaction.
We summarize this work in Sec.~\ref{sec.summary}.
\section{Theoretical Framework}
\label{sec.Theory}
\par
The muon capture reaction undergoes two steps, that is to say,
the ground-state target nucleus $ i $ is transmuted to a highly-excited state $ f $ of the daughter nucleus by the muon capture (hereafter we call this process simply the muon capture)
and
the highly-excited daughter nucleus $ f $ evaporates particles and is transmuted to the residual nucleus $ r $.
During the former process, the muon is assumed to be captured in its $ 1s $ orbital.
\par
The muon capture rate in the former process $ \omega_{fi} $ is calculated from the $ pn $-QRPA,
and the ratio of the residual nucleus $ r $ after the latter process $ P_{rf}^{\urm{emit}} $ is calculated from the HFSM.
Here, the muon wave function before the muon capture has to be considered properly.
Finally, isotope production rates by the muon capture reaction are given by
\begin{equation}
  \label{palpha}
  P_{r}
  =
  \frac{\sum_f \omega_{fi} P^{\urm{emit}}_{rf}}
  {\sum_f \omega_{fi}}.    
\end{equation}
\par
We will describe them in the following subsections.
\subsection{Muon Capture Rate}
\par
The muon capture rate is given by \cite{Walecka, Connell1972}
\begin{align}
  \omega_{fi}
  & =
    \frac{2 G^2 \nu^2}{1 + \nu/M_{\urm{T}}}
    \frac{1}{2J_i + 1}
    \notag \\
  & \times
    \sum_{M_i M_f}
    \left\{
    \sum_{JM}
    \left|
    \brakket{J_f M_f}
    {\phi_{1s} \left( \hat{\ca{M}}_{JM} - \hat{\ca{L}}_{JM} \right)}
    {J_i M_i}
    \right|^2
    \right.
    \notag \\
  & \qquad
    \left.
    +
    \sum_{JM}
    \left|
    \brakket{J_f M_f}
    {\phi_{1s} \left( \hat{\ca{T}}_{JM}^{\urm{el}} - \hat{\ca{T}}_{JM}^{\urm{mag}} \right)}
    {J_i M_i}
    \right|^2
    \right\},
    \label{eq:muoncapturerate}
\end{align}
where $ G = 1.166 \times 10^{-11} \, \mathrm{MeV}^{-2} $ \cite{CODATA2018} is the Fermi coupling constant, 
$ M_{\urm{T}} $ is the mass of the target nucleus,
$ \nu $ is the muon neutrino energy, 
and $ \phi_{1s} \equiv \phi_{1s} \left( \vec{r} \right) $ is the muon wave function of the $ 1s $ orbit. 
In this work, we restrict ourselves to study only $ \mathrm{Mo} $ isotopes with even mass number for simplicity of numerical calculation, and thus we set $ J_i = 0 $.
The definitions of 
the Coulomb and longitudinal multipole operators,
$ \hat{\ca{M}}_{JM} $ and $ \hat{\ca{L}}_{JM} $,
and
the transverse electric and magnetic multipole operators,
$ \hat{\ca{T}}_{JM}^{\urm{el}} $ and $ \hat{\ca{T}}_{JM}^{\urm{mag}} $,
in Eq.~\eqref{eq:muoncapturerate} are given in Refs.~\cite{Connell1972,Marketin2009},
where $ J $ and $ M $ satisfy
$ \vec{J}_f = \vec{J}_i + \vec{J} $ and $ M_f = M_i + M $, respectively.
From the energy conservation \cite{Marketin2009, AUERBACH1984}, 
\begin{equation}
  m_\mu + \epsilon_b +E_i 
  =
  E_f + \nu,
  \label{eq:energyconserve}
\end{equation}
where $ m_{\mu} $ is the muon mass, 
$ \epsilon_b < 0 $ is the binding energy of the muon, 
$ E_i $ and $ E_f $ are the energies of initial and final states. 
We approximate 
$ E_f - E_i = m_n - m_p + \left( \lambda_n - \lambda_p + E_{\urm{QRPA}} \right) $
\cite{Engel1999, RingandSchuck}, 
where $ m_n $ and $ m_p $ are the neutron and proton masses,
$ \lambda_n $ and $ \lambda_p $ are the neutron and proton Fermi energies of the initial nucleus,
and $ E_{\urm{QRPA}} $ is the eigenvalue of the $ pn $-QRPA equation \cite{RingandSchuck}. 
We use the effective axial-vector coupling constant $ g_{\urm{A}} = 1 $, 
instead of the free-nucleon one $ g_{\urm{A}} = 1.26 $, in the multipole operators.
\subsubsection{SHFBCS and $ pn $-QRPA}
\label{sec.pnQRPA}
\par
To calculate the transition matrix elements appeared in the curled parenthesis of Eq.~\eqref{eq:muoncapturerate}, 
the $ pn $-QRPA is used in this work.
\par
First, the ground state of the initial nucleus $ \ket{J_i M_i} $ is calculated by the SHFBCS \cite{Vautherin1973} with the SLy4 effective interaction \cite{SLy4}. 
We consider the axially-deformation of the nucleus assuming the reflection symmetry.
The mesh sizes for numerical calculations are 
$ \Delta \rho = \Delta z = 0.8 \, \mathrm{fm} $ and the box boundary conditions are $ \rho_{\urm{max}} = z_{\urm{max}} = 16 \, \mathrm{fm} $.
In the BCS approximation, the volume-type pairing force is used, 
and the neutron and proton pairing strengths are set to be 
$ V_n = 286.669 $ and $ V_p = 295.369 \, \mathrm{MeV} $ \cite{SkO'}, respectively.
The pairing active space is chosen in the same way as Ref.~\cite{SkO'}.
Under these conditions, we obtain $ \beta_2 = 0.21 $ for $ \nuc{Mo}{100}{} $ and $ \beta_2 \simeq 0.00 $ for the other $ \mathrm{Mo} $ isotopes. 
Note that although there are plenty of theoretical studies on charge-changing transitions, $\beta^+$ transition extending to a high excitation energy region as the negative muon capture is still not well established. 
In particular, time-odd components of the Skryme effective interaction, which sensitively influence spin-multipole transitions, are not understood well although important progresses have been obtained until now~\cite{Giai1981, Bender2002, Bai2011, Roca-Maza2012}. 
In this respect, this work is challenging, however we expect that we are able to obtain some hints to constraint the time-odd components of the Skyrme force from the muon capture reaction.
\par
Next, the matrix elements appeared in Eq.~\eqref{eq:muoncapturerate} are calculated by using the $ pn $-QRPA.
The $ pn $-QRPA calculation is performed by the diagonalization approach \cite{RingandSchuck}. 
The residual interaction is fully taken into account, being consistent with the ground-state calculation of the SHFBCS.
The transition matrix elements of Eq.~\eqref{eq:muoncapturerate} are thus calculated as
\begin{align}
  & \left|
    \brakket{J_f M_f}{\hat{\ca{O}}_{JM}}{J_i M_i}
    \right|^2
    \notag \\
  & =
    \left|
    \sum_{np}
    \left(
    \brakket{n}{\hat{\ca{O}}_{JM}}{\bar{p}}
    X_{np} u_n v_p
    -
    \brakket{\bar{n}}{\hat{\ca{O}}_{JM}}{p}
    Y_{np} u_p v_n
    \right)
    \right|^2.
    \label{eq:transition}
\end{align}
The ket states in Eq.~\eqref{eq:transition} of $ \ket{p} $ and $ \ket{n} $ correspond to the single-particle state of protons and neutrons, respectively, 
and $ \ket{\bar{p}} $ and $ \ket{\bar{n}} $ are their time-reversed states.
The coefficients $ X_{np} $ and $ Y_{np} $ are the forward and backward amplitudes of the $ pn $-QRPA, respectively, 
and $ u_i $ and $ v_i $ are the BCS coefficients \cite{RingandSchuck}. 
The operator $ \hat{\ca{O}}_{JM} $ is any of the multipole operator in Eq.~\eqref{eq:muoncapturerate}.
Here, we include single-particle levels up to $ 30 \, \mathrm{MeV} $ above the Fermi energies as the model space of the $ pn $-QRPA.
\par
We define the mean excitation energy of a daughter nucleus after the muon capture as
\begin{equation}
  \overline{E}
  =
  \sum_f 
  \omega_{fi}
  E_f^*,
\end{equation}
where $ E_f^* = E_{\urm{QRPA}} - E_{\urm{2qp, lowest}} $ 
and 
$ E_{\urm{2qp, lowest}} $ is the sum of the lowest proton and neutron quasiparticle energies \cite{Engel1999}.
\subsubsection{Muon wave function}
\label{sec.muon_wavefunction}
\par
The muon wave function is given by solving the Schr\"{o}dinger or Dirac equation under the external potential $ V_{\urm{pot}} $,
which is composed of two parts; 
\begin{equation}
  \label{eq:potential}
  V_{\urm{pot}} \left( r \right)
  =
  V_{\urm{$ \mu $-$ N $}} \left( r \right)
  +
  V_{\urm{$ \mu $-$ e $}} \left( r \right),
\end{equation}
where the potential is assumed to have the spherical symmetry.
\par
The former one, $ V_{\urm{$ \mu $-$ N $}} $, is the Coulomb potential due to the nucleus.
The charge distribution of the atomic nucleus $ \rho_{\urm{ch}} $ is considered in $ V_{\urm{$ \mu $-$ N $}} $ and thus it is different from the simple potential $ -Z/r $ as
\begin{equation}
  \label{eq:potential_Nmu}
  V_{\urm{$ \mu $-$ N $}} \left( r \right)
  =
  -
  4 \pi e^2
  \int_0^r
  \frac{1}{r'^2}
  \int_0^{r'}
  \overline{\rho}_{\urm{ch}} \left( r'' \right)
  r'' \, dr'' \, dr',
\end{equation}
where $ \overline{\rho}_{\urm{ch}} $ is the spherical-averaged charge distribution, that is, 
\begin{equation}
  \overline{\rho}_{\urm{ch}} \left( r \right)
  =
  \frac{1}{4 \pi}
  \int 
  \rho_{\urm{ch}} \left( \vec{r} \right)
  \, d \Omega.
\end{equation}
It should be noted that even if the nuclear charge distribution $ \rho_{\urm{ch}} $ is deformed in the intrinsic frame,
in general the spherical-averaged distribution $ \overline{\rho}_{\urm{ch}} $ should be used in Eq.~\eqref{eq:potential_Nmu},
since the muon wave function is calculated in the lab frame.
\par
To obtain the charge distribution of $ \mathrm{Mo} $ isotopes, 
the proton density calculated by the SHFBCS is convoluted with the proton form factor as follows;
\begin{equation}
  \label{Eq:charge_density}
  \rho_{\urm{ch}} \left( \vec{r} \right)
  =
  \int
  \rho_p \left( \vec{r}' \right) 
  G \left( \vec{r}' - \vec{r} \right)
  \, d \vec{r}',
\end{equation}
where the function 
$ G \left( \vec{r} \right) = \left( r_0 \sqrt{\pi} \right)^{-3} \exp \left( - \vec{r}^2 / r_0^2 \right) $ 
is the Fourier transformation of the electric form factor $ G \left( q^2 \right) $ of protons and $ \rho_p $ is the proton density distribution given by the SHFBCS.
We assume the proton root-mean-square radius 
$ \sqrt{\avr{r_p^2}} = 0.8414 \, \mathrm{fm} $ \cite{CODATA2018},
which corresponds to $ r_0 = 0.687 \, \mathrm{fm} $.
In this calculation,
first the spherical-averaged proton density distribution $ \overline{\rho}_p $ is calculated and
it is substituted into Eq.~\eqref{Eq:charge_density} to obtain $ \overline{\rho}_{\urm{ch}} $.
\par
In the practical calculation, 
to calculate Eq.~\eqref{eq:potential_Nmu}, 
the calculated spherical-averaged charge density distribution $ \overline{\rho}_{\urm{ch}} $ is fitted to the Fourier-Bessel function \cite{DeVries1987At.DataNucl.DataTables36_495},
\begin{equation}
  \overline{\rho}_{\urm{ch}} \left( r \right)
  =
  \begin{cases}
    \sum_{j = 1}^{17}
    a_j
    j_0 \left( j \pi r / R \right)
    & \text{for $ r < R $}, \\ 
    0
    & \text{for $ r > R $},
  \end{cases}
\end{equation}
where 
\begin{equation}
  j_0 \left( x \right)
  =
  \frac{\sin \left( x \right)}{x}
\end{equation}
is the spherical Bessel function.
The coefficients $ a_j $ is obtained by using \textsc{Gnuplot} 
and $ R $ is determined as the minimum value of $ r $ 
which satisfies $ \overline{\rho}_{\urm{ch}} \left( r \right) < 10^{-6} \, \mathrm{fm}^{-3} $.
\par
The latter one, $ V_{\urm{$ \mu $-$ e $}} $, is the Coulomb interaction between the muon and the electrons of the atom, which leads
\begin{equation}
  \label{eq:potential_emu}
  V_{\urm{$ \mu $-$ e $}} \left( r \right)
  =
  e^2
  \int
  \frac{\rho_e \left( \vec{r} \right)}{\left| \vec{r} - \vec{r}' \right|}
  \, d \vec{r}',
\end{equation}
where $ \rho_e $ is the electron distribution.
\par
In this work, the number of the electrons is assumed to be the same as the atomic number $ Z $, i.e., the muon is captured by the neutral atoms.
The number density of electrons $ \rho_e $ is calculated by the density functional theory (DFT) in the Dirac scheme 
\cite{Hohenberg1964Phys.Rev.136_B864, Kohn1965Phys.Rev.140_A1133, MacDonald1979J.Phys.C12_2977}
performed by the calculation package ``Atomic Density functional program PACKage (\textsc{ADPACK})'' \cite{ADPACK},
and it is assumed to be spherical symmetry.
The Perdew-Zunger exchange-correlation functional in the local density approximation,
as known as the ``PZ81'' functional \cite{Perdew1981Phys.Rev.B23_5048},
is used.
\par
After $ V_{\urm{pot}} $ is calculated,
the muon wave function is calculated numerically within the uniform mesh of $ \log r $.
\subsection{Calculation of Evaporation Residue by HFSM}
\label{code} 
\par
To describe the particle evaporation step, we used the HFSM module implemented in Comprehensive Code for Nuclear Data Evaluation (\textsc{CCONE}) developed in the Nuclear Data Center, JAEA \cite{CCONE}. 
We assume that the daughter nucleus reaches $ \ket{J_f M_f} $ the compound state soon after the muon capture and the particle evaporation follows the statistical process.
\par
Following quantities of the compound nucleus are required to run the \textsc{CCONE} as inputs:
(1) Transmission coefficients of nucleons, deuteron, triton, helium-3 are calculated by Koning-Delaroche optical potentials \cite{KandD2003} and its folding potentials. 
(2) Transmission coefficient of $ \alpha $-particle calculated from the optical potential of Avrigeanu \cite{Avrigeanu2010}, 
(3) The enhanced generalized Lorentzian function of Kopecky-Uhl \cite{KandU},
which is used for $ \gamma $-strength function,
(4) The Gilbert-Cameron method \cite{GilbertCameron} with the Mengoni-Nakajima parameter \cite{MengoniNakajima},
which is used for the nuclear level densities,
(5) Masses taken from AME2016 \cite{AME2016,AME2016b} if available and FRDM12 \cite{FRDM2012} for otherwise.
\par
It is pointed out that the contribution from the pre-equilibrium process has a non-negligible in the particle evaporation after the muon capture \cite{PhysRevC.22.2135, Hashim2018, Hashim2019}.
In addition, it can be considered that the direct process also contributes the particle emissions because the muon capture gives high energy enough to kick out protons out of nucleus directly.
However, this work considers neither the pre-equilibrium nor direct processes, which are left for our future work.
\section{Results}
\label{sec.result}
\subsection{Muon Capture Rate of $ \nuc{Mo}{\text{nat}}{} $}
\par
We first estimate the muon capture rates of $ \mathrm{Mo} $ isotopes with even mass numbers to check if our theoretical framework works well.
The calculated muon capture rates are listed in Table~\ref{tab:muoncapture} with the natural abundance ($ \mathrm{NA} $) of $ \mathrm{Mo} $ isotopes.
Only experimental data of the muon capture rate of $ \nuc{Mo}{\text{nat}}{} $ is available (see Ref.~\cite{Suzuki1987} and reference therein). 
For $ \nuc{Mo}{95}{} $ ($ \mathrm{NA} = 15.84 \, \% $) and $\nuc{Mo}{97}{}$ ($ \mathrm{NA} = 9.5 \, \% $), 
we estimate the muon capture rates by taking an average of neighboring nuclei.
The $ pn $-QRPA gives $ \omega_{fi} = 11.3 \times 10^6 \, \mathrm{s}^{-1} $ for $ \nuc{Mo}{\mathrm{nat}}{} $, which reproduces the experimental data ($ \left( 9.614 \pm 0.15 \right) \times 10^6 \, \mathrm{s}^{-1} $~\cite{Suzuki1987}) by a deviation of about $ 12 \, \% $.
Assuming that the calculated muon capture rate of $ \nuc{Mo}{100}{} $ is also overestimated by $12\%$, the expected muon capture rate is $ \omega_{fi} = 8.75 \times 10^6 \, \mathrm{s}^{-1} $.
Compared to the muon life time ($ \omega_{\urm{weak}} = 4.552 \times 10^5 \, \mathrm{s}^{-1} $ \cite{Tanabashi2018Phys.Rev.D98_030001}), the muon capture on the nucleus occurs much faster than the muon weak decay ($ \mu^- \rightarrow e^- + \overline{\nu}_e + \nu_{\mu} $).
The ratio is calculated as $ \omega_{fi} / \left( \omega_{fi} + \omega_{\urm{weak}} \right) \simeq 0.95 $.
\begin{table}[t]
  \caption{
    Calculated muon capture rates and natural abundance ($ \mathrm{NA} $) of $ \mathrm{Mo} $ isotopes with even masses. 
    Experimental muon capture rate for natural $ \mathrm{Mo} $ is also listed \cite{Suzuki1987}.}
  \label{tab:muoncapture}
  \begin{ruledtabular}
    \begin{tabular}{rdd}
      \multicolumn{1}{c}{Nucleus} & \multicolumn{1}{c}{$ \omega_{fi} $ ($ 10^{6} \, \mathrm{s}^{-1} $)} & \multicolumn{1}{c}{$ \mathrm{NA} $ ($ \mathrm{\%} $)}\\
      \hline
      $ \nuc{Mo}{92}{} $  & 13.3 & 14.53 \\
      $ \nuc{Mo}{94}{} $  & 12.2 & 9.15 \\
      $ \nuc{Mo}{96}{} $  & 11.3 & 16.67 \\
      $ \nuc{Mo}{98}{} $  & 10.3 & 24.39 \\
      $ \nuc{Mo}{100}{} $ &  9.8 & 9.82 \\
      \hline
      $ \nuc{Mo}{\text{nat}}{} $ (calc.) & 11.3 & \\
      (exp.)  & \multicolumn{1}{l}{$ 9.614 \pm 0.15 $} & \\
    \end{tabular}
  \end{ruledtabular}
\end{table}
\subsection{Muon Capture Reaction of $ \nuc{Mo}{100}{} $}
\par
\begin{figure}[t]
  \centering
  \includegraphics[width=0.9\linewidth]{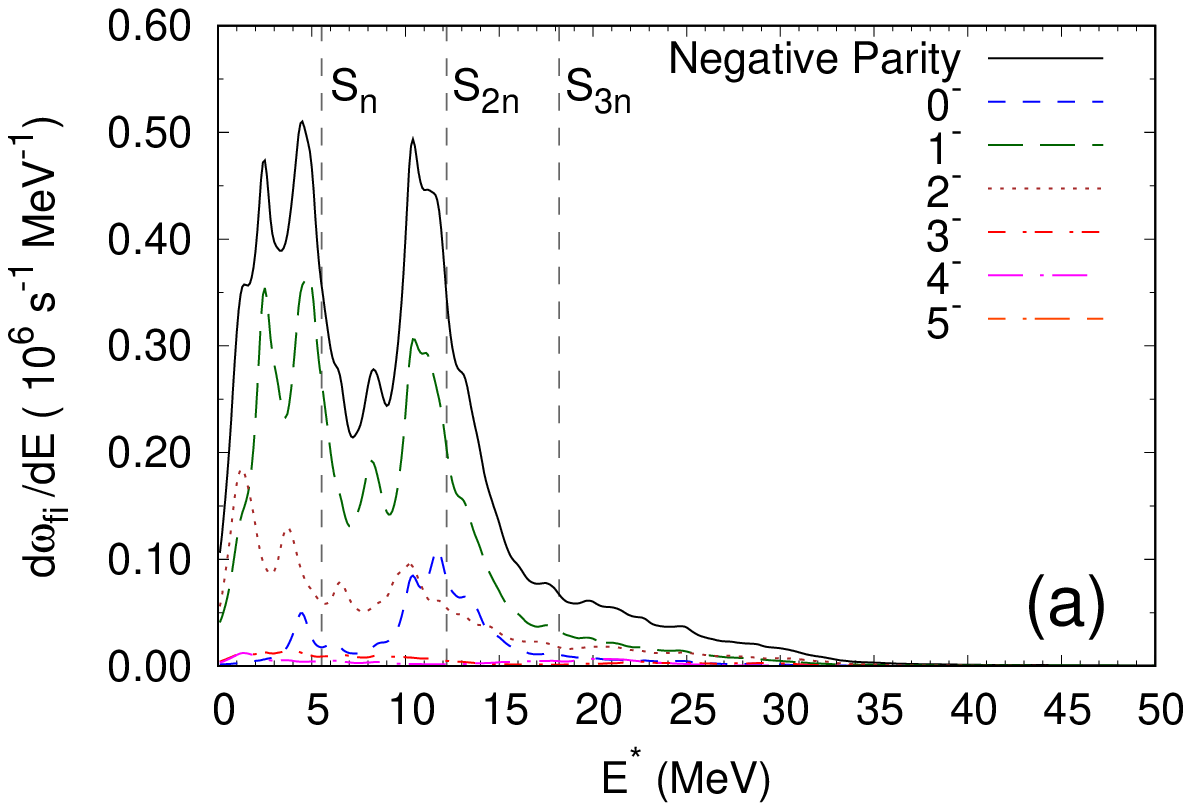}
  \includegraphics[width=0.9\linewidth]{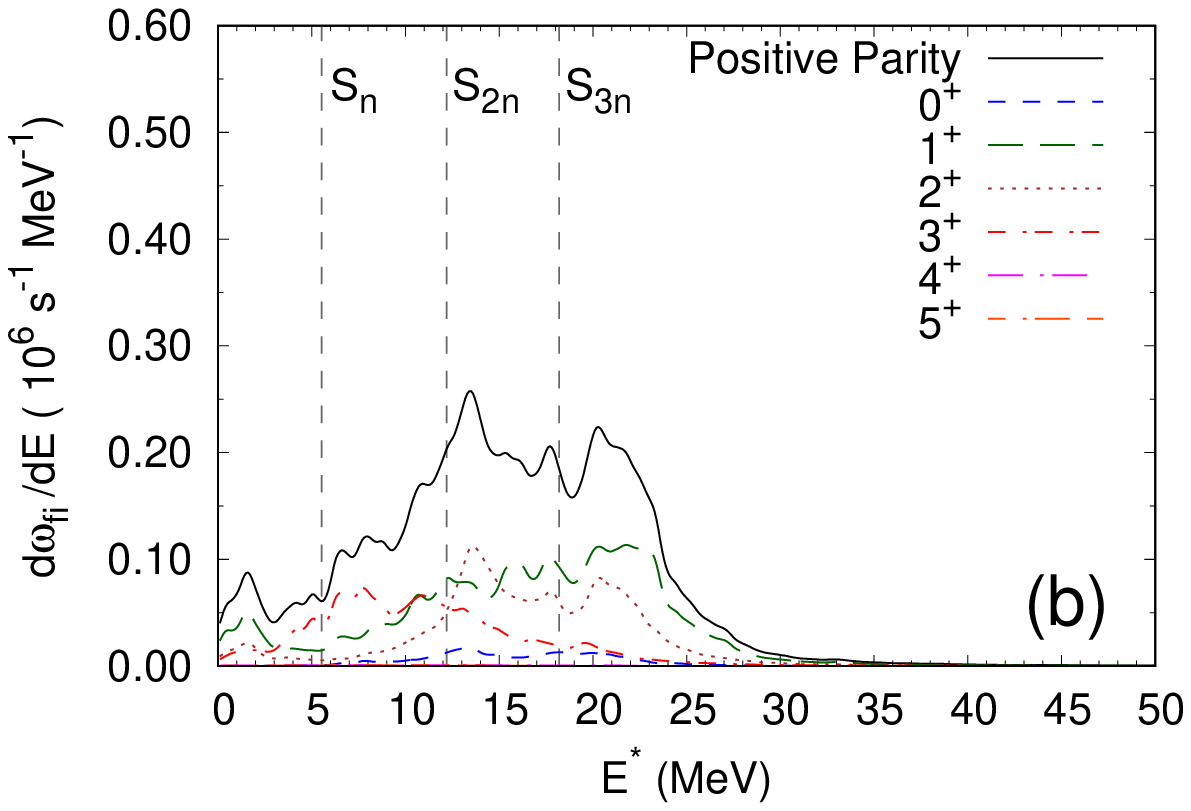}
  \caption{
    Muon capture rates as a function of excitation energy for (a) negative parity and (b) positive parity states of $ \nuc{Nb}{100}{} $. Neutron separation energies are indicated by the dotted lines.
  }
  \label{fig:excitationfunction}
\end{figure}

Figure~\ref{fig:excitationfunction} illustrates the muon capture rates of positive and negative parity states for the daughter nucleus of $\nuc{Nb}{100}{}$ as functions of excitation energy $ E^* $.
The curves shown are smoothed by a Lorentzian function with a width of $ 1 \, \mathrm{MeV} $.
We also show the separation energies of one neutron ($ S_n = 5.5 \, \mathrm{MeV} $), two neutrons ($ S_{2n} = 11.1 \, \mathrm{MeV} $), and three neutrons ($ S_{3n} = 18.4 \, \mathrm{MeV} $).
The sums of negative and positive parities are indicated by the solid lines in the panels.
We can see characteristic structures in the muon capture rates, which are different from the one estimated in Ref.~\cite{Hashim2018}.
Our result clearly indicates the importance of the nuclear structure effect on the muon capture.
The present model does not give any strong strength distributions above $ 35 \, \mathrm{MeV} $ for both negative and positive parity states.
\par
\begin{figure}[t]
  \centering
  \includegraphics[width=0.9\linewidth]{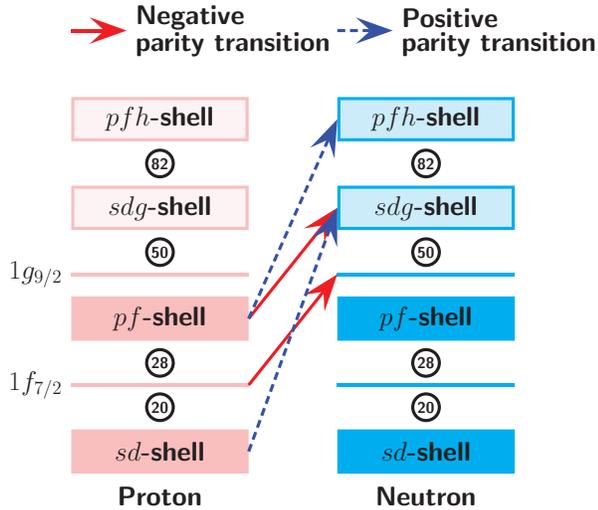}
  \caption{
    Schematic figure of transitions.
    Red-solid and blue-dashed arrows correspond to negative and positive parity transitions, respectively.
    Shells with dark color represent almost occupied states and those with light color are unoccupied ones.
    Here, the Coulomb interaction is neglected for simplicity, and thus the energies of protons and neutrons are identical.}
  \label{fig:schematic}
\end{figure}

It is clearly seen from Fig.~\ref{fig:excitationfunction} that the negative parity states distribute in low energy region and its main component spreads in $ E^* \le S_{2n} $.
On the other hand, the positive parity states distribute in higher energy region than the negative parity and its main peaks appear at above $ S_{2n} $.
This difference can be explained by considering the shell structure of $ \nuc{Nb}{100}{} $.
The numbers of nucleons of $ \nuc{Nb}{100}{} $ are close to sub-magic number $ Z = 40 $ and magic number $ N = 50 $, and thus proton and neutron $ pf $-shell and neutron intruder $ 1g $ states are almost occupied.
Therefore, the negative parity states are usually populated by a transition from proton $ pf $-shell to neutron $ sdg $-shell, so that the transition energy required is around $ 1 \hbar \omega $ if the Coulomb force is neglected.
In fact, the peaks observed in $ E^* < 5 \, \mathrm{MeV} $ for $ 1^- $ state are mainly due to the transition from proton $ 1f_{7/2} $ state to neutron $ 1g_{7/2} $ state in the spherical picture.
On the other hand, the positive parity states are mainly populated by a transition energy from proton $ sd $-shell to neutron $ sdg $-shell or proton $ pf $-shell to neutron $ pfh $-shell, so that the transition energy required is around $ 2 \hbar \omega $.
Therefore, the negative parity state is more significant than the positive parity state in a low energy region, and it is considered that the negative parity state gives large contribution to $ \nuc{Mo}{100}{} \left( \mu^-, n \right) \nuc{Nb}{99}{} $.
This mechanism is schematically shown in Fig.~\ref{fig:schematic}.
\par
From the calculated muon capture rates and excited states, particle evaporations are calculated by the HFSM.
The obtained result of $ \mathrm{Nb} $ isotope mass distribution of the $ \nuc{Mo}{100}{} \left( \mu^-, xn \right) $ is listed in Table~\ref{tab:P_alpha}.
Considering lack of our knowledge on low-lying $ \beta^+ $-type transitions and uncertainties
in spin-isospin transitions in the present theoretical model, the strength distributions of daughter nucleus yielded by the muon-capture would not be reproduced correctly. 
In spite of that, it is remarkable that the present model reasonably reproduces the production rate of the muon capture reaction.
Only for $ \nuc{Nb}{100}{} $, the production rate is rather overestimated.
This overestimation comes from too many feedings to the excited states below $ S_n $ by the muon capture and this can be observed in Fig.~\ref{fig:excitationfunction}, especially for the negative parity state.
We should keep in mind that the present theoretical model has the ambiguity as observed in Table~\ref{tab:P_alpha}.
We, however, believe that the qualitative discussion given in the following will not be affected by it.
\begin{table}[t]
  \caption{
    Production rate of $ \mathrm{Nb} $ isotopes by the muon capture reaction on $ \mathrm{Mo} $ isotopes ($ \% $). 
    The experimental data is taken from Ref.~\cite{Hashim2018}. 
    Charged particle emission rate is also listed.}
  \label{tab:P_alpha}
  \begin{ruledtabular}
    \begin{tabular}{ldd} 
      Reaction & \multicolumn{1}{c}{Experiment} & \multicolumn{1}{c}{This work} \\
      \hline
      $ \nuc{Mo}{100}{} \left( \mu^-, 0n \right)  \nuc{Nb}{100}{} $ &  8 & 28.9 \\
      $ \nuc{Mo}{100}{} \left( \mu^-, 1n \right)  \nuc{Nb}{99}{} $  & 51 & 38.1 \\
      $ \nuc{Mo}{100}{} \left( \mu^-, 2n \right)  \nuc{Nb}{98}{} $  & 16 & 23.8 \\
      $ \nuc{Mo}{100}{} \left( \mu^-, 3n \right)  \nuc{Nb}{97}{} $  & 13 & 8.73 \\
      $ \nuc{Mo}{100}{} \left( \mu^-, 4n \right)  \nuc{Nb}{96}{} $  &  6 & 0.28 \\
      $ \nuc{Mo}{100}{} \left( \mu^-, 5n \right)  \nuc{Nb}{95}{} $  &  3 & 0.01 \\ \hline
      Charged particle emission  &    & 0.06 \\
    \end{tabular}
  \end{ruledtabular}
\end{table}
\par
\begin{figure}
  \centering
  \includegraphics[width=0.98\linewidth]{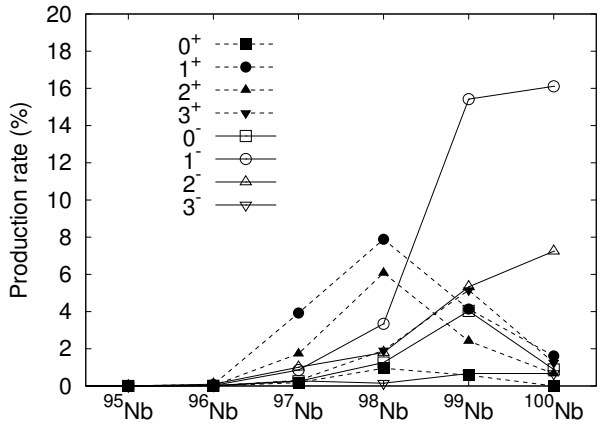}
  \caption{
    Production rate of $ \mathrm{Nb} $ isotopes by the muon capture reaction for different spin-parity states.}
  \label{fig:integrations}
\end{figure}

Now, we try to demonstrate the reason why the $ \nuc{Mo}{100}{} \left( \mu^-, n \right) \nuc{Nb}{99}{} $ reaction occurs at high probability. 
Figure \ref{fig:integrations} shows the production rate of $ \mathrm{Nb} $ isotopes by the muon capture reaction for different spin-parity states.
We did not show the contributions from $ 4^{\pm} $ and $ 5^{\pm} $ states because they are not significant.
We can clearly see that the production rate of isotopes with large mass numbers are mainly due to negative parity states and that with small mass numbers are due to positive parity states.
The production rate of $ \nuc{Nb}{99}{} $ and $ \nuc{Nb}{100}{} $ mainly comes from $ 1^- $ state as expected.
High probability of the $ \nuc{Mo}{100}{} \left( \mu^-, n \right) \nuc{Nb}{99}{} $ reaction is thus resulted from the nuclear structure of proton and neutron shells described above. 
\par
\begin{figure}
  \centering
  \includegraphics[width=0.96\linewidth]{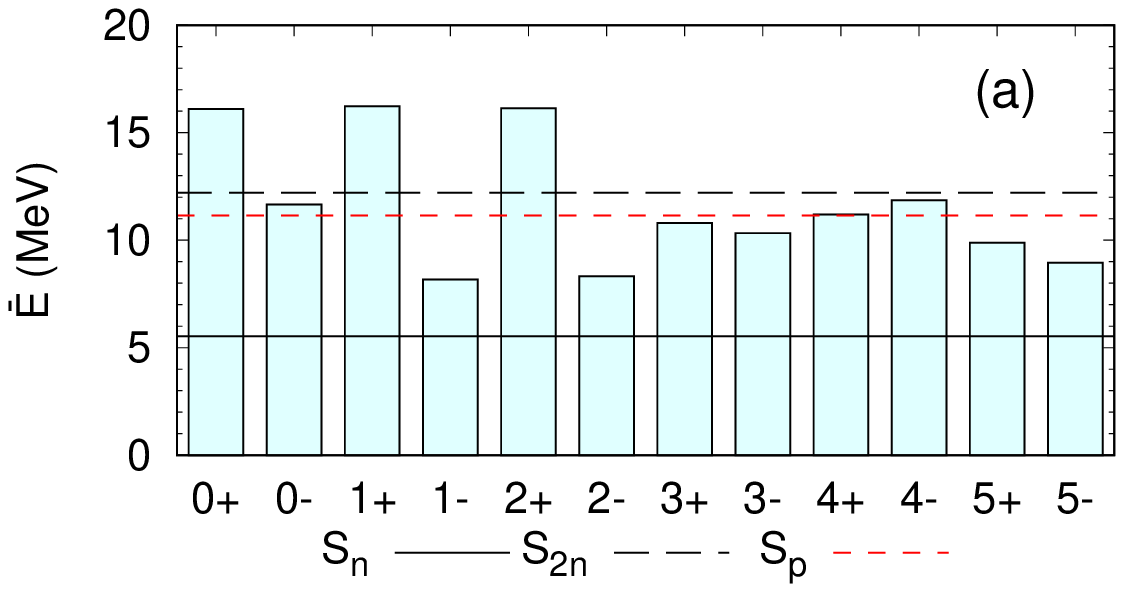} 
  \includegraphics[width=0.97\linewidth]{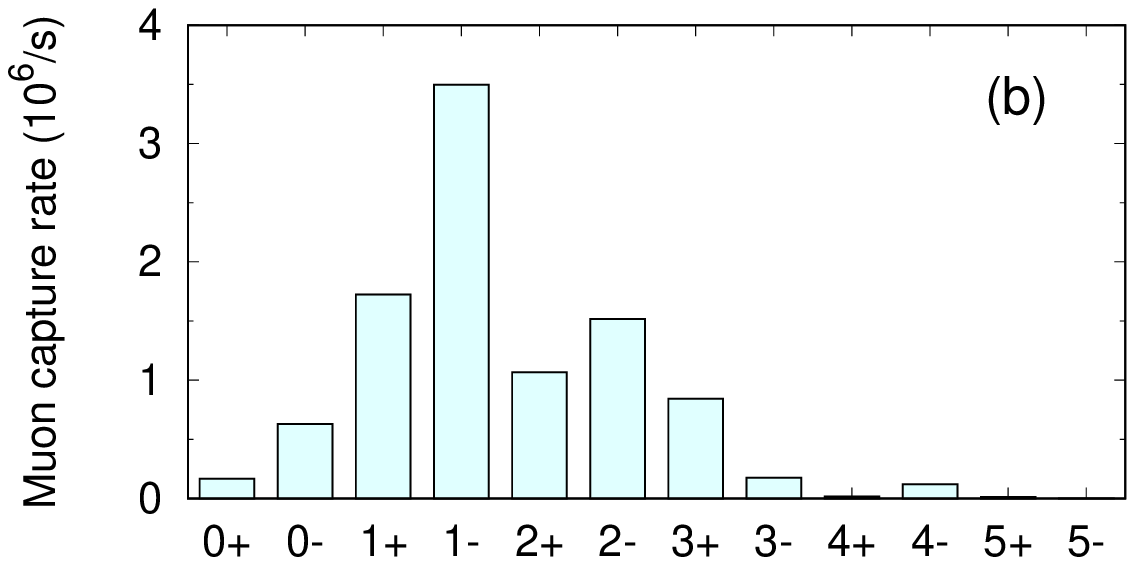}
  \caption{
    (a) Mean excitation energy $ \overline{E} $ for different spin-parity states. 
    Separation energies of one and two neutrons, and one proton for $ \nuc{Nb}{100}{} $ are also shown by the black-solid, dotted, and red-dashed lines, respectively.
    (b) Muon capture rates on $ \nuc{Mo}{100}{} $ for different spin-parity states of $ \nuc{Nb}{100}{} $.}
  \label{fig:ave_Energy}
\end{figure}

To qualitatively understand the isotope production rates of the muon capture reaction, we illustrate $ \overline{E} $ and the muon capture rates for different spin-parity states in Fig.~\ref{fig:ave_Energy}. 
The dashed, dotted, and long-dashed lines in the panel (a) are $ S_n $, $ S_{2n} $, and one-proton separation energies ($ S_p = 9.5 \, \mathrm{MeV} $) of $ \nuc{Nb}{100}{} $. 
The mean excitation energies of every state exceed the one neutron separation energy, and $ \nuc{Nb}{99}{} $ is easy to be produced.
Among $ 0^{\pm} $, $ 1 ^{\pm} $, and $ 2^{\pm} $ states, 
the positive parity states have a larger mean excitation energy than the negative parity states as expected,
and the mean excitation energies exceed $ S_{2n} $ for $ 0^+ $, $ 1^+ $, and $ 2^+ $ states.
\par
The panel (b) of Fig.~\ref{fig:ave_Energy} shows the muon capture rate $ \omega_{fi} $
of different spin-parity states of $ \nuc{Nb}{100}{} $.
The $ 1^- $ and $ 2^- $ states have the largest and the third largest contributions.
Their $ \overline{E} $ is below $ S_{2n} $ and $ S_p $,
and thus they easily evaporate one neutron and are transmuted to $ \nuc{Nb}{99}{} $.
The second largest contribution is $ 1^+ $ state with $ \overline{E} \simeq 16 \, \mathrm{MeV} $, 
which is greater than $ S_{2n} $ and $ S_p $,
and thus it is transmuted to $ \nuc{Nb}{99}{} $ as well as $ \nuc{Nb}{98}{} $ and $ \nuc{Zr}{100}{} $.
The muon capture rates of $ 0^+ $ and $ 2^+ $ states are relatively small,
which contributes $ \nuc{Nb}{99}{} $ production less.
As a result, high probability of $ \nuc{Mo}{100}{} \left( \mu^-, n \right) \nuc{No}{99}{} $ reaction can be explained by the mean excitation energy.
\par
From Table~\ref{tab:P_alpha}, the rate of charged particle emission by $ \nuc{Mo}{100}{} \left( \mu^-, x \right) $ is only $ 0.06 \, \% $. 
The hindrance of the charged particle emissions is simply understood from the mean excitation energy.
As seen in Fig.~\ref{fig:ave_Energy}, the mean excitation energies are greater than $ S_p $ only for $ 0^{\pm} $, $ 1^+ $, $ 2^+ $, and $ 4^- $ states. 
Even though the excitation energies of those states are higher than $ S_p $, 
the neutron emission is easier to occur than proton emission because $ S_n $ is $ 4 \, \mathrm{MeV} $ lower than $ S_p $.
Emitted neutron withdraws energy and the compound nucleus is no longer able to emit protons.
The Coulomb barrier also hinders the proton emission, as well.
For $ \alpha $-particle emission, $ Q_{\alpha} \simeq 3.1 \, \mathrm{MeV} $ is lower than $ S_n $.
However, it has a larger Coulomb barrier than proton, and transmission probability is thus expected to be small. 
\par
At last of this subsection, the hindrance of the charged particle emission is discussed among $ \mathrm{Mo} $ isotopes.
Because $\nuc{Mo}{100}{}$ locates at a relatively neutron-rich side in the nuclear chart, $ S_p $ is much higher than other $ \mathrm{Mo} $ isotopes.
Therefore, the charged particle emission may occur at higher probability for $ \mathrm{Mo} $ isotopes with small mass numbers than $ \nuc{Mo}{100}{} $.
To investigate them, we calculate the isotope production rates by the muon capture reaction on other $ \mathrm{Mo} $ isotopes, and the results are shown in Table~\ref{tab:P_alpha2}. 
The reaction rates of $ \left( \mu^-, xn \right) $ are similar to $ \nuc{Mo}{100}{} $ case.
In any isotope, one neutron emission occurs at high probability as the muon capture reaction on $ \nuc{Mo}{100}{} $.
In contrast, the charged particle emission rate for the muon capture reaction varies with respect to the mass number.
As decreasing the mass number, the charged particle emission rate becomes larger,
since the neutron and proton separation energies are 
$ S_n = 7.2 $ and $ S_p = 6.5 \, \mathrm{MeV} $ for $ \nuc{Nb}{94}{} $, 
$ S_n = 6.9 $ and $ S_p = 7.2 \, \mathrm{MeV} $ for $ \nuc{Nb}{96}{} $, 
and $ S_n = 6.0 $ and $ S_p = 7.9 \, \mathrm{MeV} $ for $ \nuc{Nb}{98}{} $.
Finally, the rate on $ \nuc{Nb}{92}{} $ gives a high probability of about $ 13 \, \% $.
This is because the proton separation energy of $ \nuc{Nb}{92}{} $ ($ S_p = 5.8 \, \mathrm{MeV} $) is lower than the neutron separation energy ($ S_n = 7.9 \, \mathrm{MeV} $),
and thus the proton emission occurs much easier in $ \nuc{Nb}{92}{} $ than in $ \nuc{Nb}{100}{} $.
\begin{table}[t]
  \caption{
    Production of $ \mathrm{Nb} $ isotopes by the muon capture reaction on $ \mathrm{Mo} $ isotopes ($ \% $). 
    Charged particle emission rate is also listed.}
  \label{tab:P_alpha2}
  \begin{ruledtabular}
    \begin{tabular}{cd} 
      Reaction & \multicolumn{1}{c}{This work} \\
      \hline
      $ \nuc{Mo}{92}{} \left( \mu^-, 0n \right) \nuc{Nb}{92}{} $ & 34.3 \\
      $ \nuc{Mo}{92}{} \left( \mu^-, 1n \right) \nuc{Nb}{91}{} $ & 43.0 \\
      $ \nuc{Mo}{92}{} \left( \mu^-, 2n \right) \nuc{Nb}{90}{} $ & 9.48 \\
      $ \nuc{Mo}{92}{} \left( \mu^-, 3n \right) \nuc{Nb}{89}{} $ & 0.12 \\
      $ \nuc{Mo}{92}{} \left( \mu^-, 4n \right) \nuc{Nb}{88}{} $ & 0.00 \\
      $ \nuc{Mo}{92}{} \left( \mu^-, 5n \right) \nuc{Nb}{87}{} $ & 0.00 \\  
      Charged particle emission & 13.1 \\  
      \hline
      $ \nuc{Mo}{94}{} \left( \mu^-, 0n \right) \nuc{Nb}{94}{} $ & 33.2 \\
      $ \nuc{Mo}{94}{} \left( \mu^-, 1n \right) \nuc{Nb}{93}{} $ & 43.4 \\
      $ \nuc{Mo}{94}{} \left( \mu^-, 2n \right) \nuc{Nb}{92}{} $ & 20.1 \\
      $ \nuc{Mo}{94}{} \left( \mu^-, 3n \right) \nuc{Nb}{91}{} $ & 2.27 \\
      $ \nuc{Mo}{94}{} \left( \mu^-, 4n \right) \nuc{Nb}{90}{} $ & 0.01 \\
      $ \nuc{Mo}{94}{} \left( \mu^-, 5n \right) \nuc{Nb}{89}{} $ & 0.00 \\  
      Charged particle emission & 1.12 \\  
      \hline
      $ \nuc{Mo}{96}{} \left( \mu^-, 0n \right) \nuc{Nb}{96}{} $ & 32.1 \\
      $ \nuc{Mo}{96}{} \left( \mu^-, 1n \right) \nuc{Nb}{95}{} $ & 43.3 \\
      $ \nuc{Mo}{96}{} \left( \mu^-, 2n \right) \nuc{Nb}{94}{} $ & 20.8 \\
      $ \nuc{Mo}{96}{} \left( \mu^-, 3n \right) \nuc{Nb}{93}{} $ & 3.54 \\
      $ \nuc{Mo}{96}{} \left( \mu^-, 4n \right) \nuc{Nb}{92}{} $ & 0.07 \\
      $ \nuc{Mo}{96}{} \left( \mu^-, 5n \right) \nuc{Nb}{91}{} $ & 0.00 \\  
      Charged particle emission & 0.22 \\  
      \hline
      $ \nuc{Mo}{98}{} \left( \mu^-, 0n \right) \nuc{Nb}{98}{} $ & 31.0 \\
      $ \nuc{Mo}{98}{} \left( \mu^-, 1n \right) \nuc{Nb}{97}{} $ & 43.5 \\
      $ \nuc{Mo}{98}{} \left( \mu^-, 2n \right) \nuc{Nb}{96}{} $ & 20.9 \\
      $ \nuc{Mo}{98}{} \left( \mu^-, 3n \right) \nuc{Nb}{95}{} $ & 4.43 \\
      $ \nuc{Mo}{98}{} \left( \mu^-, 4n \right) \nuc{Nb}{94}{} $ & 0.11 \\
      $ \nuc{Mo}{98}{} \left( \mu^-, 5n \right) \nuc{Nb}{93}{} $ & 0.00 \\  
      Charged particle emission & 0.05 \\  
    \end{tabular}
  \end{ruledtabular}
\end{table}
\section{Feasibility of $ \nuc{Mo}{99}{} $ production by muon capture}
\label{sec.economical}
\par
To produce $ \nuc{Mo}{99}{} $ by the muon capture, high flux of negative muon beam is required. 
Basically, the negative muon is produced by the following reaction sequence:
\begin{equation}
  \label{Eq:muonproduction}
  \begin{aligned}
    p + N
    & \to
    \pi^- + N', \\
    \pi^-
    & \to
    \mu^- + \overline{\nu}_{\mu}.
  \end{aligned}
\end{equation}
Currently, there are several facilities providing negative muon sources in the world, 
which are ISIS Neutron and Muon Source in RAL, MUSE in J-PARC, and MuSIC in RCNP, Osaka Univ.
New facilities or beam lines for high-intensity muon flux are also scheduled in COMET, J-PARC (Japan) and Mu2e, FNAL (USA) and EMus, CSNS (China).
We will show the production of $ \nuc{Mo}{99}{} $ through the muon capture reaction on these facilities.
\begin{table*}[t]
  \centering 
  \caption{Comparison of profiles of muon beam facilities of MUSE, MuSIC, and MERIT, and the number of $ \nuc{Mo}{99}{} $ expected in those facilities. 
    The isotope production through $ \left( n, 2n \right) $ reaction is also listed.}
  \label{comp}
  \begin{ruledtabular}
    \begin{tabular}{lrddr|rr} 
      Facility & \multicolumn{1}{l}{Reaction for} & \multicolumn{1}{r}{Particle Energy} & \multicolumn{1}{r}{Electric Power} & \multicolumn{1}{r|}{Muon Flux $ I_\mu $} & \multicolumn{2}{c}{Number of $ \nuc{Mo}{99}{} $} \\
               & \multicolumn{1}{l}{secondary beam} & \multicolumn{1}{r}{($ \mathrm{MeV} $)} & \multicolumn{1}{r}{($ \mathrm{kW} $)} & \multicolumn{1}{r|}{($ s^{-1} $)} & \multicolumn{1}{r}{($ s^{-1} $)} & \multicolumn{1}{r}{($ s^{-1} \, \mathrm{kW}^{-1} $)} \\
      \hline
      MUSE, J-PARC \cite{J-parc} & $ C \left( p, \pi^- \right) $ 
                                                  & 3000 & 1000 & $ \sim 1 \times 10^6 $ & $ \sim 4.8 \times 10^5 $ & $ \sim 4.8 \times 10^2 $ \\ 
      MuSIC, RCNP \cite{Hashim2019} & $ C \left( p, \pi^- \right) $ 
                                                  & 400 & 0.4 & $ 4.4 \times 10^5 $ & $ 2.1 \times 10^4 $ & $ 5.3 \times 10^4 $ \\
                                                  % 
                                                  % EMuS, China \cite{EMus}     & C($p$,$\pi^-$)  & $ 1600 $ & 500 & $1\times10^{8}$ & $5.0\times10^{7}$ & $1.0\times10^{5}$ \\
                                                  % 
                                                  % COMET, J-PARC \cite{COMET}     & C($p$,$\pi^-$)  & $ 8000 $ & 50 & $1\times10^{11}$ & $5.0\times10^{10}$ & $1.0\times10^{9}$ \\
                                                  % 
      MERIT \cite{Merit}     & $ \mathrm{Li} \left( p, \pi^- \right) $ 
                                                  & 800 & 1600 & $ 1 \times 10^{16} $ & $ 4.8 \times 10^{15} $ & $ 3.0 \times 10^{12} $ \\
      \hline
      $ \nuc{Mo}{100}{} \left( n, 2n \right) \nuc{Mo}{99}{} $ \cite{Minato2017} & $ \nuc{C}{\text{nat}}{} \left( d, xn \right) $ 
                                                  & 40 & 80 & --- & $ 2.6 \times 10^9 $ & $ 3.3 \times 10^{7} $ \\
    \end{tabular}
  \end{ruledtabular}
  \label{table:facilities}
\end{table*}
\par
Let us estimate the $ \nuc{Mo}{99}{} $ production rate by the muon capture reaction.
We assumed that $ 100 \, \% $ of the incident muons are captured by an orbit of $ \nuc{Mo}{100}{} $ and forms a muonic atom.
Then, the production rate of $ \nuc{Nb}{99}{} $ reads
\begin{equation}
  \label{eq:productionrate}
  R_{\urm{$ \mathrm{Nb} $-99}}
  =
  I_{\mu}
  \frac{\omega_{fi}}{\omega_{fi} + \omega_{\urm{weak}}}
  P_{\urm{$ \mathrm{Nb} $-99}},
\end{equation}
where $ I_\mu $ is the muon flux available at experimental facilities
and $ P_{\urm{$ \mathrm{Nb} $-99}} $ is the muon capture rate in Eq.~\eqref{eq:muoncapturerate}.
We here neglect efficiencies arising from technical difficulties such as muon transport efficiency and so on.
\par
The results of the production rate 
(in unit of $ \mathrm{s}^{-1} $ and $ \mathrm{s}^{-1} \, \mathrm{kW}^{-1} $) 
are shown together in Table~\ref{table:facilities}.
We picked up two muon beam facilities, which are the MUSE and the MuSIC, and the profiles are also shown in the table.
We also list the production rate of $ \nuc{Mo}{100}{} \left( n, 2n \right) \nuc{Mo}{99}{} $ \cite{Nagai2009} (hereafter $ \left( n,2n \right) $).
The number of $ \nuc{Mo}{99}{} $ obtained by the muon capture is about $ 4.8 \times 10^5 $ and $ 2.1 \times 10^4 \, \mathrm{s}^{-1} $ for the MUSE and the MuSIC, respectively.
Compared to $ \left( n, 2n \right) $, one of the most efficient production method, the production rate of $ \nuc{Mo}{99}{} $ through the muon capture reaction is low.
Looking at the result of production rate per electric power deposited, $ \left( n, 2n \right) $ reaction is more efficient than the muon capture.
Therefore, much higher flux of negative muon source is required to reach an efficient production of $ \nuc{Mo}{99}{} $ through the muon capture reaction.
\par
Abe \textit{et al\/.} also studied RI productions by the muon capture reaction and assessed its feasibility for nuclear transmutation \cite{Abe2016}, using PHITS code \cite{Sato2018}.
They also concluded that a high intensity of negative muon source is required for transmutation by negative muon capture reaction.
However, any of the facility mentioned above is not designed for negative muon productions.
It would be also important to discuss the production rate with a facility optimized to produce negative muon sources.
Recently, a breakthrough idea to produce negative muon, called the multiplex energy recovery internal target (MERIT), was proposed \cite{Merit}.
The MERIT is able to recycle proton beams passing through the target every time. 
The protons are re-accelerated and be stored in the MERIT ring.
With this idea, $ 1 \times 10^{16} \, \mathrm{s}^{-1} $ of muon flux can be obtained in the condition of $ 2 \, \mathrm{mA} $ of $ 800 \, \mathrm{MeV} $ deuteron \cite{Merit}.
If assuming that the MERIT can provide ideal muon flux in accordance with its blueprint, the number of $ \nuc{Mo}{99}{} $ amounts to $ 4.8 \times 10^{15} \, \mathrm{s}^{-1} $, which is six order times larger than the RI production through $ \left( n, 2n \right) $ reaction. 
Comparing the result in terms of electric power, $ \nuc{Mo}{99}{} $ production using the MERIT is about five order of magnitude more efficient than $ \left( n, 2n \right) $ reaction.
Note that, however, there still remains technical issues in the $ \nuc{Mo}{99}{} $ production by the muon capture reaction with MERIT.
There would be actually some barriers in terms of technical issues until one accomplishes an efficient $ \nuc{Mo}{99}{} $ production and supply being comparable to other methods.
\section{Summary}
\label{sec.summary}
\par
In order to understand the detailed mechanism of $ \mathrm{Nb} $ isotope production through the muon capture reaction on $ \nuc{Mo}{100}{} $, 
we studied the muon capture and subsequent particle evaporation with a microscopic theoretical model.
We used the $ pn $-QRPA on the basis of SLy4 energy density functional for the muon capture and the Hauser-Feshbach statistical model for the particle evaporation process.
\par
Our framework gives a reasonable agreement with the experimental data of the muon capture rate on $ \nuc{Mo}{\text{nat}}{} $. 
From the calculation, it is found that negative parity states populated by the muon capture on $ \nuc{Mo}{100}{} $ have a major contribution to the muon capture reaction at low excitation energy, while positive parity states have at higher excitation energy.
We demonstrated this difference by considering the nuclear shell structure of $ \nuc{Nb}{100}{} $.
\par
Our framework reasonably reproduced the experimentally measured $ \mathrm{Nb} $ isotope production.
A high probability of $ \nuc{Mo}{100}{} \left( \mu^-,n \right) \nuc{Nb}{99}{} $ could be explained by the $ 1^- $ state populated by the muon capture, which has a large contribution to the muon capture reaction at excitation energies around $ S_n $.
This finding and the hindrance of charged particle emissions are also discussed qualitatively by using mean excitation energy.
\par
We also study the feasibility of the $ \nuc{Mo}{99}{} $ isotope production through the muon capture reaction by three different muon beam sources. 
The $ \nuc{Mo}{99}{} $ isotope production by the muon capture reaction is less efficient than $ \left( n, 2n \right) $ production method.
However, if high flux muon beam, as proposed as the MERIT, is obtained, the muon capture reaction can be a promising method for $ \nuc{Mo}{99}{} $ production as well as other RI productions.
\par
While there are plenty of theoretical studies on charge-changing transitions, 
$ \beta^+ $ transition extending to a high excitation energy region as the negative muon capture causes is not studied well. 
In addition, the muon capture involves high multipole transitions, including natural and unnatural parities.
To understand those physics, further effort is demanded both from experimental and theoretical sides, and the muon capture reaction, especially the isotope production rate, may provide us an important insight for improvement of theoretical models.
In our model theoretical framework, pre-equilibrium and direct processes are omitted. 
Those contributions would improve the results obtained in isotope production.
We plan to include those contributions in our framework as a first development.
\begin{acknowledgments}
  The authors acknowledge S.~Abe (JAEA) for his valuable cooperation about PHITS calculations.
  TN would like to thank the RIKEN iTHEMS program,
  the JSPS-NSFC Bilateral Program for Joint Research Project on Nuclear mass and life for unravelling mysteries of the $ r $-process,
  and the JSPS Grant-in-Aid for JSPS Fellows under Grant No.~19J20543.
  Some numerical calculations have been performed on cluster computers at the RIKEN iTHEMS program.
\end{acknowledgments}
\appendix
\section{Derivation of $ V_{\urm{$ N $-$ \mu $}} $}
\par
In this appendix, the derivation of $ V_{\urm{$ N $-$ \mu $}} $ under the spherical-averaged charge distribution $ \overline{\rho}_{\urm{ch}} $ is given.
In the appendices, the dielectric constant of vacuum, $ \epsilon_0 $, is shown explicitly.
\par
According to the Maxwell equation \cite{Greiner1998ClassicalElectrodynamics_Springer-Verlag}, 
the charge distribution forms the electric field $ \ve{E}_N $;
\begin{equation}
  \label{eq:Maxwell_div}
  \epsilon_0 \ve{E}_N \left( r \right)
  =
  e^2
  \Div \overline{\rho}_{\urm{ch}} \left( r \right) ,
\end{equation}
where $ \ve{E}_N $ also holds the spherical symmetry.
Equation \eqref{eq:Maxwell_div} can be rewritten as
\begin{equation}
  \label{eq:Maxwell_int}
  4 \pi r^2 E_N \left( r \right)
  =
  \frac{4 \pi e^2}{\epsilon_0}
  \int_0^r 
  \overline{\rho}_{\urm{ch}} \left( r' \right) \,
  r'^2
  \, dr'.
\end{equation}
Since a potential formed by $ \overline{\rho}_{\urm{ch}} $ satisfies
\begin{equation}
  E_N \left( r \right)
  =
  - \frac{d V_{\urm{$ N $-$ \mu $}} \left( r \right)}{dr},
\end{equation}
the potential reads
\begin{align}
  V_{\urm{$ N $-$ \mu $}} \left( r \right)
  & =
    - \int_0^r 
    E_N \left( r' \right) \, dr'
    \notag \\
  & =
    -
    \frac{e^2}{\epsilon_0}
    \int_0^r
    \frac{1}{r'^2}
    \int_0^{r'}
    \overline{\rho}_{\urm{ch}} \left( r'' \right) \,
    r''^2 \, dr''.
\end{align}
In the unit we use in this paper $ \epsilon_0 = 1/4 \pi $ is hold and thus Eq.~\eqref{eq:potential_Nmu} is given.
\section{Calculable expression $ V_{\urm{$ e $-$ \mu $}} $ under the spherical symmetry}
\par
In this appendix, an efficiently calculable form of Eq.~\eqref{eq:potential_emu} is given.
The total charge inside the sphere with radius $ r $ due to $ \rho_e $ is written as
\begin{equation}
  \label{eq:totE}
  Q \left( r \right)
  =
  4 \pi e
  \int_0^r
  \rho_e \left( r \right) \,
  r^2 \, dr
\end{equation}
and the electric field formed by $ \rho_e $ at $ r $ is
\begin{equation}
  E_e \left( r \right)
  =
  \frac{e}{4 \pi \epsilon_0}
  \frac{Q \left( r \right)}{r^2}.
\end{equation}
Therefore, the potential due to $ \rho_e $ is
\begin{align}
  V_{\urm{$ e $-$ \mu $}} \left( r \right)
  & =
    e
    \int_r^{\infty}
    E_e \left( r' \right) \, dr'
    \notag \\
  & =
    \frac{e^2}{4 \pi \epsilon_0}
    \int_r^{\infty}
    \frac{Q \left( r' \right)}{r'^2}
    \, dr'
    \notag \\
  & =
    \frac{e^2}{4 \pi \epsilon_0}
    \int_r^{\infty}
    \left[
    \frac{d}{dr'}
    \left( - \frac{1}{r'} \right)
    \right]
    Q \left( r' \right)
    \, d r'
    \notag \\
  & =
    \frac{e^2}{4 \pi \epsilon_0}
    \frac{Q \left( r \right)}{r}
    +
    \frac{e^2}{4 \pi \epsilon_0}
    \int_r^{\infty}
    \frac{1}{r'}
    \frac{d Q \left( r' \right)}{dr'}
    \, dr'
    \notag \\
  & =
    \frac{e^2}{\epsilon_0 r}
    \int_0^r
    \rho_e \left( r' \right) \, 
    r'^2 \, dr'
    +
    \frac{e^2}{\epsilon_0}
    \int_r^{\infty}
    \rho_e \left( r' \right) \, 
    r' \, dr'.
\end{align}
In the unit we use in this paper $ \epsilon_0 = 1/4 \pi $ is hold and thus 
\begin{equation}
  V_{\urm{$ e $-$ \mu $}} \left( r \right)
  = 
  \frac{4 \pi e^2}{r}
  \int_0^r
  \rho_e \left( r' \right) \, 
  r'^2 \, dr'
  +
  4 \pi e^2 
  \int_r^{\infty}
  \rho_e \left( r' \right) \, 
  r' \, dr'
\end{equation}
is given.
\par
Note that the similar calculation method is already used for the Hartree term in the ADPACK \cite{ADPACK}.
% \clearpage
\bibliography{biblio}

\begin{thebibliography}{51}%
\makeatletter
\providecommand \@ifxundefined [1]{%
 \@ifx{#1\undefined}
}%
\providecommand \@ifnum [1]{%
 \ifnum #1\expandafter \@firstoftwo
 \else \expandafter \@secondoftwo
 \fi
}%
\providecommand \@ifx [1]{%
 \ifx #1\expandafter \@firstoftwo
 \else \expandafter \@secondoftwo
 \fi
}%
\providecommand \natexlab [1]{#1}%
\providecommand \enquote  [1]{``#1''}%
\providecommand \bibnamefont  [1]{#1}%
\providecommand \bibfnamefont [1]{#1}%
\providecommand \citenamefont [1]{#1}%
\providecommand \href@noop [0]{\@secondoftwo}%
\providecommand \href [0]{\begingroup \@sanitize@url \@href}%
\providecommand \@href[1]{\@@startlink{#1}\@@href}%
\providecommand \@@href[1]{\endgroup#1\@@endlink}%
\providecommand \@sanitize@url [0]{\catcode `\\12\catcode `\$12\catcode
  `\&12\catcode `\#12\catcode `\^12\catcode `\_12\catcode `\%12\relax}%
\providecommand \@@startlink[1]{}%
\providecommand \@@endlink[0]{}%
\providecommand \url  [0]{\begingroup\@sanitize@url \@url }%
\providecommand \@url [1]{\endgroup\@href {#1}{\urlprefix }}%
\providecommand \urlprefix  [0]{URL }%
\providecommand \Eprint [0]{\href }%
\providecommand \doibase [0]{http://dx.doi.org/}%
\providecommand \selectlanguage [0]{\@gobble}%
\providecommand \bibinfo  [0]{\@secondoftwo}%
\providecommand \bibfield  [0]{\@secondoftwo}%
\providecommand \translation [1]{[#1]}%
\providecommand \BibitemOpen [0]{}%
\providecommand \bibitemStop [0]{}%
\providecommand \bibitemNoStop [0]{.\EOS\space}%
\providecommand \EOS [0]{\spacefactor3000\relax}%
\providecommand \BibitemShut  [1]{\csname bibitem#1\endcsname}%
\let\auto@bib@innerbib\@empty
%</preamble>
\bibitem [{\citenamefont {Van~Noorden}(2013)}]{Noorden2013}%
  \BibitemOpen
  \bibfield  {author} {\bibinfo {author} {\bibfnamefont {R.}~\bibnamefont
  {Van~Noorden}},\ }\href {\doibase 10.1038/504202a} {\bibfield  {journal}
  {\bibinfo  {journal} {Nature}\ }\textbf {\bibinfo {volume} {504}},\ \bibinfo
  {pages} {202} (\bibinfo {year} {2013})}\BibitemShut {NoStop}%
\bibitem [{\citenamefont {Cherry}\ \emph {et~al.}(2003)\citenamefont {Cherry},
  \citenamefont {Sorenson},\ and\ \citenamefont {Phelps}}]{Cherry2003}%
  \BibitemOpen
  \bibfield  {author} {\bibinfo {author} {\bibfnamefont {S.~R.}\ \bibnamefont
  {Cherry}}, \bibinfo {author} {\bibfnamefont {J.~A.}\ \bibnamefont
  {Sorenson}}, \ and\ \bibinfo {author} {\bibfnamefont {M.~E.}\ \bibnamefont
  {Phelps}},\ }\href@noop {} {\emph {\bibinfo {title} {{Physics in Nuclear
  Medicine}}}}\ (\bibinfo  {publisher} {Saunders},\ \bibinfo {address}
  {Philadelphia, PA},\ \bibinfo {year} {2003})\BibitemShut {NoStop}%
\bibitem [{\citenamefont {Nagai}\ and\ \citenamefont
  {Hatsukawa}(2009)}]{Nagai2009}%
  \BibitemOpen
  \bibfield  {author} {\bibinfo {author} {\bibfnamefont {Y.}~\bibnamefont
  {Nagai}}\ and\ \bibinfo {author} {\bibfnamefont {Y.}~\bibnamefont
  {Hatsukawa}},\ }\href {\doibase 10.1143/JPSJ.78.033201} {\bibfield  {journal}
  {\bibinfo  {journal} {J. Phys. Soc. Jpn.}\ }\textbf {\bibinfo {volume}
  {78}},\ \bibinfo {pages} {033201} (\bibinfo {year} {2009})}\BibitemShut
  {NoStop}%
\bibitem [{\citenamefont {Nagai}\ \emph {et~al.}(2013)\citenamefont {Nagai},
  \citenamefont {Hashimoto}, \citenamefont {Hatsukawa}, \citenamefont {Saeki},
  \citenamefont {Motoishi}, \citenamefont {Sato}, \citenamefont {Kawabata},
  \citenamefont {Harada}, \citenamefont {Kin}, \citenamefont {Tsukada},
  \citenamefont {K.~Sato}, \citenamefont {Minato}, \citenamefont {Iwamoto},
  \citenamefont {Iwamoto}, \citenamefont {Seki}, \citenamefont {Yokoyama},
  \citenamefont {Shiina}, \citenamefont {Ohta}, \citenamefont {Takeuchi},
  \citenamefont {Kawauchi}, \citenamefont {Sato}, \citenamefont {Yamabayashi},
  \citenamefont {Adachi}, \citenamefont {Kikuchi}, \citenamefont {Mitsumoto},\
  and\ \citenamefont {Igarashi}}]{Nagai2013}%
  \BibitemOpen
  \bibfield  {author} {\bibinfo {author} {\bibfnamefont {Y.}~\bibnamefont
  {Nagai}}, \bibinfo {author} {\bibfnamefont {K.}~\bibnamefont {Hashimoto}},
  \bibinfo {author} {\bibfnamefont {Y.}~\bibnamefont {Hatsukawa}}, \bibinfo
  {author} {\bibfnamefont {H.}~\bibnamefont {Saeki}}, \bibinfo {author}
  {\bibfnamefont {S.}~\bibnamefont {Motoishi}}, \bibinfo {author}
  {\bibfnamefont {N.}~\bibnamefont {Sato}}, \bibinfo {author} {\bibfnamefont
  {M.}~\bibnamefont {Kawabata}}, \bibinfo {author} {\bibfnamefont
  {H.}~\bibnamefont {Harada}}, \bibinfo {author} {\bibfnamefont
  {T.}~\bibnamefont {Kin}}, \bibinfo {author} {\bibfnamefont {K.}~\bibnamefont
  {Tsukada}}, \bibinfo {author} {\bibfnamefont {T.}~\bibnamefont {K.~Sato}},
  \bibinfo {author} {\bibfnamefont {F.}~\bibnamefont {Minato}}, \bibinfo
  {author} {\bibfnamefont {O.}~\bibnamefont {Iwamoto}}, \bibinfo {author}
  {\bibfnamefont {N.}~\bibnamefont {Iwamoto}}, \bibinfo {author} {\bibfnamefont
  {Y.}~\bibnamefont {Seki}}, \bibinfo {author} {\bibfnamefont {K.}~\bibnamefont
  {Yokoyama}}, \bibinfo {author} {\bibfnamefont {T.}~\bibnamefont {Shiina}},
  \bibinfo {author} {\bibfnamefont {A.}~\bibnamefont {Ohta}}, \bibinfo {author}
  {\bibfnamefont {N.}~\bibnamefont {Takeuchi}}, \bibinfo {author}
  {\bibfnamefont {Y.}~\bibnamefont {Kawauchi}}, \bibinfo {author}
  {\bibfnamefont {N.}~\bibnamefont {Sato}}, \bibinfo {author} {\bibfnamefont
  {H.}~\bibnamefont {Yamabayashi}}, \bibinfo {author} {\bibfnamefont
  {Y.}~\bibnamefont {Adachi}}, \bibinfo {author} {\bibfnamefont
  {Y.}~\bibnamefont {Kikuchi}}, \bibinfo {author} {\bibfnamefont
  {T.}~\bibnamefont {Mitsumoto}}, \ and\ \bibinfo {author} {\bibfnamefont
  {T.}~\bibnamefont {Igarashi}},\ }\href {\doibase 10.7566/JPSJ.82.064201}
  {\bibfield  {journal} {\bibinfo  {journal} {J. Phys. Soc. Jpn.}\ }\textbf
  {\bibinfo {volume} {82}},\ \bibinfo {pages} {064201} (\bibinfo {year}
  {2013})}\BibitemShut {NoStop}%
\bibitem [{\citenamefont {Minato}\ \emph {et~al.}(2017)\citenamefont {Minato},
  \citenamefont {Tsukada}, \citenamefont {Sato}, \citenamefont {Watanabe},
  \citenamefont {Saeki}, \citenamefont {Kawabata}, \citenamefont {Hashimoto},\
  and\ \citenamefont {Nagai}}]{Minato2017}%
  \BibitemOpen
  \bibfield  {author} {\bibinfo {author} {\bibfnamefont {F.}~\bibnamefont
  {Minato}}, \bibinfo {author} {\bibfnamefont {K.}~\bibnamefont {Tsukada}},
  \bibinfo {author} {\bibfnamefont {N.}~\bibnamefont {Sato}}, \bibinfo {author}
  {\bibfnamefont {S.}~\bibnamefont {Watanabe}}, \bibinfo {author}
  {\bibfnamefont {H.}~\bibnamefont {Saeki}}, \bibinfo {author} {\bibfnamefont
  {M.}~\bibnamefont {Kawabata}}, \bibinfo {author} {\bibfnamefont
  {S.}~\bibnamefont {Hashimoto}}, \ and\ \bibinfo {author} {\bibfnamefont
  {Y.}~\bibnamefont {Nagai}},\ }\href {\doibase 10.7566/JPSJ.86.114803}
  {\bibfield  {journal} {\bibinfo  {journal} {J. Phys. Soc. Jpn.}\ }\textbf
  {\bibinfo {volume} {86}},\ \bibinfo {pages} {114803} (\bibinfo {year}
  {2017})}\BibitemShut {NoStop}%
\bibitem [{\citenamefont {Nagamine}(2003)}]{Nagamine2003}%
  \BibitemOpen
  \bibfield  {author} {\bibinfo {author} {\bibfnamefont {K.}~\bibnamefont
  {Nagamine}},\ }\href {\doibase 10.1017/CBO9780511470776} {\emph {\bibinfo
  {title} {Introductory Muon Science}}}\ (\bibinfo  {publisher} {Cambridge
  University Press},\ \bibinfo {year} {2003})\BibitemShut {NoStop}%
\bibitem [{\citenamefont {Hashim}\ \emph {et~al.}(2018)\citenamefont {Hashim},
  \citenamefont {Ejiri}, \citenamefont {Shima}, \citenamefont {Takahisa},
  \citenamefont {Sato}, \citenamefont {Kuno}, \citenamefont {Ninomiya},
  \citenamefont {Kawamura},\ and\ \citenamefont {Miyake}}]{Hashim2018}%
  \BibitemOpen
  \bibfield  {author} {\bibinfo {author} {\bibfnamefont {I.~H.}\ \bibnamefont
  {Hashim}}, \bibinfo {author} {\bibfnamefont {H.}~\bibnamefont {Ejiri}},
  \bibinfo {author} {\bibfnamefont {T.}~\bibnamefont {Shima}}, \bibinfo
  {author} {\bibfnamefont {K.}~\bibnamefont {Takahisa}}, \bibinfo {author}
  {\bibfnamefont {A.}~\bibnamefont {Sato}}, \bibinfo {author} {\bibfnamefont
  {Y.}~\bibnamefont {Kuno}}, \bibinfo {author} {\bibfnamefont {K.}~\bibnamefont
  {Ninomiya}}, \bibinfo {author} {\bibfnamefont {N.}~\bibnamefont {Kawamura}},
  \ and\ \bibinfo {author} {\bibfnamefont {Y.}~\bibnamefont {Miyake}},\ }\href
  {\doibase 10.1103/PhysRevC.97.014617} {\bibfield  {journal} {\bibinfo
  {journal} {Phys. Rev. C}\ }\textbf {\bibinfo {volume} {97}},\ \bibinfo
  {pages} {014617} (\bibinfo {year} {2018})}\BibitemShut {NoStop}%
\bibitem [{\citenamefont {Hashim}\ \emph {et~al.}(2019)\citenamefont {Hashim},
  \citenamefont {Ejiri}, \citenamefont {Shima}, \citenamefont {Takahisa},
  \citenamefont {Sato}, \citenamefont {Kuno}, \citenamefont {Ninomiya},
  \citenamefont {Kawamura},\ and\ \citenamefont {Miyake}}]{Hashim2019}%
  \BibitemOpen
  \bibfield  {author} {\bibinfo {author} {\bibfnamefont {I.}~\bibnamefont
  {Hashim}}, \bibinfo {author} {\bibfnamefont {H.}~\bibnamefont {Ejiri}},
  \bibinfo {author} {\bibfnamefont {T.}~\bibnamefont {Shima}}, \bibinfo
  {author} {\bibfnamefont {K.}~\bibnamefont {Takahisa}}, \bibinfo {author}
  {\bibfnamefont {A.}~\bibnamefont {Sato}}, \bibinfo {author} {\bibfnamefont
  {Y.}~\bibnamefont {Kuno}}, \bibinfo {author} {\bibfnamefont {K.}~\bibnamefont
  {Ninomiya}}, \bibinfo {author} {\bibfnamefont {N.}~\bibnamefont {Kawamura}},
  \ and\ \bibinfo {author} {\bibfnamefont {Y.}~\bibnamefont {Miyake}},\
  }\href@noop {} {\  (\bibinfo {year} {2019})},\ \Eprint
  {http://arxiv.org/abs/1908.08166} {arXiv:1908.08166 [nucl-ex]} \BibitemShut
  {NoStop}%
\bibitem [{\citenamefont {Ejiri}\ and\ \citenamefont
  {de~Voight}(1989)}]{Ejiri1989}%
  \BibitemOpen
  \bibfield  {author} {\bibinfo {author} {\bibfnamefont {H.}~\bibnamefont
  {Ejiri}}\ and\ \bibinfo {author} {\bibfnamefont {M.~J.~A.}\ \bibnamefont
  {de~Voight}},\ }\href@noop {} {\emph {\bibinfo {title} {{Gamma ray and
  electron spectroscopy in nuclear physics}}}}\ (\bibinfo  {publisher} {Oxford
  University Press},\ \bibinfo {year} {1989})\BibitemShut {NoStop}%
\bibitem [{\citenamefont {Halbleib}\ and\ \citenamefont
  {Sorensen}(1967)}]{Halbleib1967Nucl.Phys.A98_542}%
  \BibitemOpen
  \bibfield  {author} {\bibinfo {author} {\bibfnamefont {J.~A.}\ \bibnamefont
  {Halbleib}}\ and\ \bibinfo {author} {\bibfnamefont {R.~A.}\ \bibnamefont
  {Sorensen}},\ }\href {\doibase 10.1016/0375-9474(67)90098-X} {\bibfield
  {journal} {\bibinfo  {journal} {Nucl. Phys. A}\ }\textbf {\bibinfo {volume}
  {98}},\ \bibinfo {pages} {542} (\bibinfo {year} {1967})}\BibitemShut
  {NoStop}%
\bibitem [{\citenamefont {Vautherin}(1973)}]{Vautherin1973}%
  \BibitemOpen
  \bibfield  {author} {\bibinfo {author} {\bibfnamefont {D.}~\bibnamefont
  {Vautherin}},\ }\href {\doibase 10.1103/PhysRevC.7.296} {\bibfield  {journal}
  {\bibinfo  {journal} {Phys. Rev. C}\ }\textbf {\bibinfo {volume} {7}},\
  \bibinfo {pages} {296} (\bibinfo {year} {1973})}\BibitemShut {NoStop}%
\bibitem [{\citenamefont {Hauser}\ and\ \citenamefont {Feshbach}(1952)}]{HFSM}%
  \BibitemOpen
  \bibfield  {author} {\bibinfo {author} {\bibfnamefont {W.}~\bibnamefont
  {Hauser}}\ and\ \bibinfo {author} {\bibfnamefont {H.}~\bibnamefont
  {Feshbach}},\ }\href {\doibase 10.1103/PhysRev.87.366} {\bibfield  {journal}
  {\bibinfo  {journal} {Phys. Rev.}\ }\textbf {\bibinfo {volume} {87}},\
  \bibinfo {pages} {366} (\bibinfo {year} {1952})}\BibitemShut {NoStop}%
\bibitem [{\citenamefont {Kolbe}\ \emph {et~al.}(1994)\citenamefont {Kolbe},
  \citenamefont {Langanke},\ and\ \citenamefont {Vogel}}]{Kolbe1994}%
  \BibitemOpen
  \bibfield  {author} {\bibinfo {author} {\bibfnamefont {E.}~\bibnamefont
  {Kolbe}}, \bibinfo {author} {\bibfnamefont {K.}~\bibnamefont {Langanke}}, \
  and\ \bibinfo {author} {\bibfnamefont {P.}~\bibnamefont {Vogel}},\ }\href
  {\doibase 10.1103/PhysRevC.50.2576} {\bibfield  {journal} {\bibinfo
  {journal} {Phys. Rev. C}\ }\textbf {\bibinfo {volume} {50}},\ \bibinfo
  {pages} {2576} (\bibinfo {year} {1994})}\BibitemShut {NoStop}%
\bibitem [{\citenamefont {Kolbe}\ \emph {et~al.}(2000)\citenamefont {Kolbe},
  \citenamefont {Langanke},\ and\ \citenamefont {Vogel}}]{Kolbe2000}%
  \BibitemOpen
  \bibfield  {author} {\bibinfo {author} {\bibfnamefont {E.}~\bibnamefont
  {Kolbe}}, \bibinfo {author} {\bibfnamefont {K.}~\bibnamefont {Langanke}}, \
  and\ \bibinfo {author} {\bibfnamefont {P.}~\bibnamefont {Vogel}},\ }\href
  {\doibase 10.1103/PhysRevC.62.055502} {\bibfield  {journal} {\bibinfo
  {journal} {Phys. Rev. C}\ }\textbf {\bibinfo {volume} {62}},\ \bibinfo
  {pages} {055502} (\bibinfo {year} {2000})}\BibitemShut {NoStop}%
\bibitem [{\citenamefont {Zinner}\ \emph {et~al.}(2006)\citenamefont {Zinner},
  \citenamefont {Langanke},\ and\ \citenamefont {Vogel}}]{Zinner2006}%
  \BibitemOpen
  \bibfield  {author} {\bibinfo {author} {\bibfnamefont {N.~T.}\ \bibnamefont
  {Zinner}}, \bibinfo {author} {\bibfnamefont {K.}~\bibnamefont {Langanke}}, \
  and\ \bibinfo {author} {\bibfnamefont {P.}~\bibnamefont {Vogel}},\ }\href
  {\doibase 10.1103/PhysRevC.74.024326} {\bibfield  {journal} {\bibinfo
  {journal} {Phys. Rev. C}\ }\textbf {\bibinfo {volume} {74}},\ \bibinfo
  {pages} {024326} (\bibinfo {year} {2006})}\BibitemShut {NoStop}%
\bibitem [{\citenamefont {Marketin}\ \emph {et~al.}(2009)\citenamefont
  {Marketin}, \citenamefont {Paar}, \citenamefont {Nik\v{s}i\'{c}},\ and\
  \citenamefont {Vretenar}}]{Marketin2009}%
  \BibitemOpen
  \bibfield  {author} {\bibinfo {author} {\bibfnamefont {T.}~\bibnamefont
  {Marketin}}, \bibinfo {author} {\bibfnamefont {N.}~\bibnamefont {Paar}},
  \bibinfo {author} {\bibfnamefont {T.}~\bibnamefont {Nik\v{s}i\'{c}}}, \ and\
  \bibinfo {author} {\bibfnamefont {D.}~\bibnamefont {Vretenar}},\ }\href
  {\doibase 10.1103/PhysRevC.79.054323} {\bibfield  {journal} {\bibinfo
  {journal} {Phys. Rev. C}\ }\textbf {\bibinfo {volume} {79}},\ \bibinfo
  {pages} {054323} (\bibinfo {year} {2009})}\BibitemShut {NoStop}%
\bibitem [{\citenamefont {Hughes}\ and\ \citenamefont {Wu}(1975)}]{Walecka}%
  \BibitemOpen
  \bibinfo {editor} {\bibfnamefont {V.~W.}\ \bibnamefont {Hughes}}\ and\
  \bibinfo {editor} {\bibfnamefont {C.}~\bibnamefont {Wu}},\ eds.,\ \href
  {\doibase https://doi.org/10.1016/B978-0-12-360602-0.X5001-5} {\emph
  {\bibinfo {title} {{Muon Physics}}}}\ (\bibinfo  {publisher} {Academic
  Press},\ \bibinfo {address} {New York},\ \bibinfo {year} {1975})\BibitemShut
  {NoStop}%
\bibitem [{\citenamefont {O'Connell}\ \emph {et~al.}(1972)\citenamefont
  {O'Connell}, \citenamefont {Donnelly},\ and\ \citenamefont
  {Walecka}}]{Connell1972}%
  \BibitemOpen
  \bibfield  {author} {\bibinfo {author} {\bibfnamefont {J.~S.}\ \bibnamefont
  {O'Connell}}, \bibinfo {author} {\bibfnamefont {T.~W.}\ \bibnamefont
  {Donnelly}}, \ and\ \bibinfo {author} {\bibfnamefont {J.~D.}\ \bibnamefont
  {Walecka}},\ }\href {\doibase 10.1103/PhysRevC.6.719} {\bibfield  {journal}
  {\bibinfo  {journal} {Phys. Rev. C}\ }\textbf {\bibinfo {volume} {6}},\
  \bibinfo {pages} {719} (\bibinfo {year} {1972})}\BibitemShut {NoStop}%
\bibitem [{\citenamefont {Tiesinga}\ \emph {et~al.}(2019)\citenamefont
  {Tiesinga}, \citenamefont {Mohr}, \citenamefont {Newell},\ and\ \citenamefont
  {Taylor}}]{CODATA2018}%
  \BibitemOpen
  \bibfield  {author} {\bibinfo {author} {\bibfnamefont {E.}~\bibnamefont
  {Tiesinga}}, \bibinfo {author} {\bibfnamefont {P.~J.}\ \bibnamefont {Mohr}},
  \bibinfo {author} {\bibfnamefont {D.~B.}\ \bibnamefont {Newell}}, \ and\
  \bibinfo {author} {\bibfnamefont {B.~N.}\ \bibnamefont {Taylor}},\ }\href
  {http://physics.nist.gov/constants} {\enquote {\bibinfo {title} {{The 2018
  CODATA Recommended Values of the Fundamental Physical Constants (Web Version
  8.0).}}}\ } (\bibinfo {year} {2019}),\ \bibinfo {note} {mD 20899}\BibitemShut
  {NoStop}%
\bibitem [{\citenamefont {Auerbach}\ and\ \citenamefont
  {Klein}(1984)}]{AUERBACH1984}%
  \BibitemOpen
  \bibfield  {author} {\bibinfo {author} {\bibfnamefont {N.}~\bibnamefont
  {Auerbach}}\ and\ \bibinfo {author} {\bibfnamefont {A.}~\bibnamefont
  {Klein}},\ }\href {\doibase 10.1016/0375-9474(84)90360-9} {\bibfield
  {journal} {\bibinfo  {journal} {Nucl. Phys. A}\ }\textbf {\bibinfo {volume}
  {422}},\ \bibinfo {pages} {480} (\bibinfo {year} {1984})}\BibitemShut
  {NoStop}%
\bibitem [{\citenamefont {Engel}\ \emph {et~al.}(1999)\citenamefont {Engel},
  \citenamefont {Bender}, \citenamefont {Dobaczewski}, \citenamefont
  {Nazarewicz},\ and\ \citenamefont {Surman}}]{Engel1999}%
  \BibitemOpen
  \bibfield  {author} {\bibinfo {author} {\bibfnamefont {J.}~\bibnamefont
  {Engel}}, \bibinfo {author} {\bibfnamefont {M.}~\bibnamefont {Bender}},
  \bibinfo {author} {\bibfnamefont {J.}~\bibnamefont {Dobaczewski}}, \bibinfo
  {author} {\bibfnamefont {W.}~\bibnamefont {Nazarewicz}}, \ and\ \bibinfo
  {author} {\bibfnamefont {R.}~\bibnamefont {Surman}},\ }\href {\doibase
  10.1103/PhysRevC.60.014302} {\bibfield  {journal} {\bibinfo  {journal} {Phys.
  Rev. C}\ }\textbf {\bibinfo {volume} {60}},\ \bibinfo {pages} {014302}
  (\bibinfo {year} {1999})}\BibitemShut {NoStop}%
\bibitem [{\citenamefont {Ring}\ and\ \citenamefont
  {Schuck}(1980)}]{RingandSchuck}%
  \BibitemOpen
  \bibfield  {author} {\bibinfo {author} {\bibfnamefont {P.}~\bibnamefont
  {Ring}}\ and\ \bibinfo {author} {\bibfnamefont {P.}~\bibnamefont {Schuck}},\
  }\href@noop {} {\emph {\bibinfo {title} {{The Nuclear Many-Body Problem}}}}\
  (\bibinfo  {publisher} {Springer-Verlag},\ \bibinfo {address} {Berlin},\
  \bibinfo {year} {1980})\BibitemShut {NoStop}%
\bibitem [{\citenamefont {Chabanat}\ \emph {et~al.}(1998)\citenamefont
  {Chabanat}, \citenamefont {Bonche}, \citenamefont {Haensel}, \citenamefont
  {Meyer},\ and\ \citenamefont {Schaeffer}}]{SLy4}%
  \BibitemOpen
  \bibfield  {author} {\bibinfo {author} {\bibfnamefont {E.}~\bibnamefont
  {Chabanat}}, \bibinfo {author} {\bibfnamefont {P.}~\bibnamefont {Bonche}},
  \bibinfo {author} {\bibfnamefont {P.}~\bibnamefont {Haensel}}, \bibinfo
  {author} {\bibfnamefont {J.}~\bibnamefont {Meyer}}, \ and\ \bibinfo {author}
  {\bibfnamefont {R.}~\bibnamefont {Schaeffer}},\ }\href {\doibase
  10.1016/S0375-9474(98)00180-8} {\bibfield  {journal} {\bibinfo  {journal}
  {Nucl. Phys. A}\ }\textbf {\bibinfo {volume} {635}},\ \bibinfo {pages} {231}
  (\bibinfo {year} {1998})}\BibitemShut {NoStop}%
\bibitem [{\citenamefont {Reinhard}\ \emph {et~al.}(1999)\citenamefont
  {Reinhard}, \citenamefont {Dean}, \citenamefont {Nazarewicz}, \citenamefont
  {Dobaczewski}, \citenamefont {Maruhn},\ and\ \citenamefont {Strayer}}]{SkO'}%
  \BibitemOpen
  \bibfield  {author} {\bibinfo {author} {\bibfnamefont {P.-G.}\ \bibnamefont
  {Reinhard}}, \bibinfo {author} {\bibfnamefont {D.~J.}\ \bibnamefont {Dean}},
  \bibinfo {author} {\bibfnamefont {W.}~\bibnamefont {Nazarewicz}}, \bibinfo
  {author} {\bibfnamefont {J.}~\bibnamefont {Dobaczewski}}, \bibinfo {author}
  {\bibfnamefont {J.~A.}\ \bibnamefont {Maruhn}}, \ and\ \bibinfo {author}
  {\bibfnamefont {M.~R.}\ \bibnamefont {Strayer}},\ }\href {\doibase
  10.1103/PhysRevC.60.014316} {\bibfield  {journal} {\bibinfo  {journal} {Phys.
  Rev. C}\ }\textbf {\bibinfo {volume} {60}},\ \bibinfo {pages} {014316}
  (\bibinfo {year} {1999})}\BibitemShut {NoStop}%
\bibitem [{\citenamefont {Giai}\ and\ \citenamefont {Sagawa}(1981)}]{Giai1981}%
  \BibitemOpen
  \bibfield  {author} {\bibinfo {author} {\bibfnamefont {N.~V.}\ \bibnamefont
  {Giai}}\ and\ \bibinfo {author} {\bibfnamefont {H.}~\bibnamefont {Sagawa}},\
  }\href {\doibase https://doi.org/10.1016/0370-2693(81)90646-8} {\bibfield
  {journal} {\bibinfo  {journal} {Physics Letters B}\ }\textbf {\bibinfo
  {volume} {106}},\ \bibinfo {pages} {379 } (\bibinfo {year}
  {1981})}\BibitemShut {NoStop}%
\bibitem [{\citenamefont {Bender}\ \emph {et~al.}(2002)\citenamefont {Bender},
  \citenamefont {Dobaczewski}, \citenamefont {Engel},\ and\ \citenamefont
  {Nazarewicz}}]{Bender2002}%
  \BibitemOpen
  \bibfield  {author} {\bibinfo {author} {\bibfnamefont {M.}~\bibnamefont
  {Bender}}, \bibinfo {author} {\bibfnamefont {J.}~\bibnamefont {Dobaczewski}},
  \bibinfo {author} {\bibfnamefont {J.}~\bibnamefont {Engel}}, \ and\ \bibinfo
  {author} {\bibfnamefont {W.}~\bibnamefont {Nazarewicz}},\ }\href {\doibase
  10.1103/PhysRevC.65.054322} {\bibfield  {journal} {\bibinfo  {journal} {Phys.
  Rev. C}\ }\textbf {\bibinfo {volume} {65}},\ \bibinfo {pages} {054322}
  (\bibinfo {year} {2002})}\BibitemShut {NoStop}%
\bibitem [{\citenamefont {Bai}\ \emph {et~al.}(2011)\citenamefont {Bai},
  \citenamefont {Zhang}, \citenamefont {Sagawa}, \citenamefont {Zhang},
  \citenamefont {Col\`o},\ and\ \citenamefont {Xu}}]{Bai2011}%
  \BibitemOpen
  \bibfield  {author} {\bibinfo {author} {\bibfnamefont {C.~L.}\ \bibnamefont
  {Bai}}, \bibinfo {author} {\bibfnamefont {H.~Q.}\ \bibnamefont {Zhang}},
  \bibinfo {author} {\bibfnamefont {H.}~\bibnamefont {Sagawa}}, \bibinfo
  {author} {\bibfnamefont {X.~Z.}\ \bibnamefont {Zhang}}, \bibinfo {author}
  {\bibfnamefont {G.}~\bibnamefont {Col\`o}}, \ and\ \bibinfo {author}
  {\bibfnamefont {F.~R.}\ \bibnamefont {Xu}},\ }\href {\doibase
  10.1103/PhysRevC.83.054316} {\bibfield  {journal} {\bibinfo  {journal} {Phys.
  Rev. C}\ }\textbf {\bibinfo {volume} {83}},\ \bibinfo {pages} {054316}
  (\bibinfo {year} {2011})}\BibitemShut {NoStop}%
\bibitem [{\citenamefont {Roca-Maza}\ \emph {et~al.}(2012)\citenamefont
  {Roca-Maza}, \citenamefont {Col\`o},\ and\ \citenamefont
  {Sagawa}}]{Roca-Maza2012}%
  \BibitemOpen
  \bibfield  {author} {\bibinfo {author} {\bibfnamefont {X.}~\bibnamefont
  {Roca-Maza}}, \bibinfo {author} {\bibfnamefont {G.}~\bibnamefont {Col\`o}}, \
  and\ \bibinfo {author} {\bibfnamefont {H.}~\bibnamefont {Sagawa}},\ }\href
  {\doibase 10.1103/PhysRevC.86.031306} {\bibfield  {journal} {\bibinfo
  {journal} {Phys. Rev. C}\ }\textbf {\bibinfo {volume} {86}},\ \bibinfo
  {pages} {031306} (\bibinfo {year} {2012})}\BibitemShut {NoStop}%
\bibitem [{\citenamefont {De~Vries}\ \emph {et~al.}(1987)\citenamefont
  {De~Vries}, \citenamefont {De~Jager},\ and\ \citenamefont
  {De~Vries}}]{DeVries1987At.DataNucl.DataTables36_495}%
  \BibitemOpen
  \bibfield  {author} {\bibinfo {author} {\bibfnamefont {H.}~\bibnamefont
  {De~Vries}}, \bibinfo {author} {\bibfnamefont {C.}~\bibnamefont {De~Jager}},
  \ and\ \bibinfo {author} {\bibfnamefont {C.}~\bibnamefont {De~Vries}},\
  }\href {\doibase 10.1016/0092-640X(87)90013-1} {\bibfield  {journal}
  {\bibinfo  {journal} {At. Data Nucl. Data Tables}\ }\textbf {\bibinfo
  {volume} {36}},\ \bibinfo {pages} {495} (\bibinfo {year} {1987})}\BibitemShut
  {NoStop}%
\bibitem [{\citenamefont {Hohenberg}\ and\ \citenamefont
  {Kohn}(1964)}]{Hohenberg1964Phys.Rev.136_B864}%
  \BibitemOpen
  \bibfield  {author} {\bibinfo {author} {\bibfnamefont {P.}~\bibnamefont
  {Hohenberg}}\ and\ \bibinfo {author} {\bibfnamefont {W.}~\bibnamefont
  {Kohn}},\ }\href {\doibase 10.1103/PhysRev.136.B864} {\bibfield  {journal}
  {\bibinfo  {journal} {Phys. Rev.}\ }\textbf {\bibinfo {volume} {136}},\
  \bibinfo {pages} {B864} (\bibinfo {year} {1964})}\BibitemShut {NoStop}%
\bibitem [{\citenamefont {Kohn}\ and\ \citenamefont
  {Sham}(1965)}]{Kohn1965Phys.Rev.140_A1133}%
  \BibitemOpen
  \bibfield  {author} {\bibinfo {author} {\bibfnamefont {W.}~\bibnamefont
  {Kohn}}\ and\ \bibinfo {author} {\bibfnamefont {L.~J.}\ \bibnamefont
  {Sham}},\ }\href {\doibase 10.1103/PhysRev.140.A1133} {\bibfield  {journal}
  {\bibinfo  {journal} {Phys. Rev.}\ }\textbf {\bibinfo {volume} {140}},\
  \bibinfo {pages} {A1133} (\bibinfo {year} {1965})}\BibitemShut {NoStop}%
\bibitem [{\citenamefont {MacDonald}\ and\ \citenamefont
  {Vosko}(1979)}]{MacDonald1979J.Phys.C12_2977}%
  \BibitemOpen
  \bibfield  {author} {\bibinfo {author} {\bibfnamefont {A.~H.}\ \bibnamefont
  {MacDonald}}\ and\ \bibinfo {author} {\bibfnamefont {S.~H.}\ \bibnamefont
  {Vosko}},\ }\href {\doibase 10.1088/0022-3719/12/15/007} {\bibfield
  {journal} {\bibinfo  {journal} {J. Phys. C}\ }\textbf {\bibinfo {volume}
  {12}},\ \bibinfo {pages} {2977} (\bibinfo {year} {1979})}\BibitemShut
  {NoStop}%
\bibitem [{\citenamefont {Ozaki}\ \emph {et~al.}(2011)\citenamefont {Ozaki},
  \citenamefont {Kino}, \citenamefont {Kawai},\ and\ \citenamefont
  {Toyoda}}]{ADPACK}%
  \BibitemOpen
  \bibfield  {author} {\bibinfo {author} {\bibfnamefont {T.}~\bibnamefont
  {Ozaki}}, \bibinfo {author} {\bibfnamefont {H.}~\bibnamefont {Kino}},
  \bibinfo {author} {\bibfnamefont {H.}~\bibnamefont {Kawai}}, \ and\ \bibinfo
  {author} {\bibfnamefont {M.}~\bibnamefont {Toyoda}},\ }\href@noop {}
  {\enquote {\bibinfo {title} {{ADPACK Ver.2.2}},}\ }\bibinfo {howpublished}
  {\url{http://www.openmx-square.org/adpack_man2.2/}} (\bibinfo {year}
  {2011})\BibitemShut {NoStop}%
\bibitem [{\citenamefont {Perdew}\ and\ \citenamefont
  {Zunger}(1981)}]{Perdew1981Phys.Rev.B23_5048}%
  \BibitemOpen
  \bibfield  {author} {\bibinfo {author} {\bibfnamefont {J.~P.}\ \bibnamefont
  {Perdew}}\ and\ \bibinfo {author} {\bibfnamefont {A.}~\bibnamefont
  {Zunger}},\ }\href {\doibase 10.1103/PhysRevB.23.5048} {\bibfield  {journal}
  {\bibinfo  {journal} {Phys. Rev. B}\ }\textbf {\bibinfo {volume} {23}},\
  \bibinfo {pages} {5048} (\bibinfo {year} {1981})}\BibitemShut {NoStop}%
\bibitem [{\citenamefont {Iwamoto}\ \emph {et~al.}(2016)\citenamefont
  {Iwamoto}, \citenamefont {Iwamoto}, \citenamefont {Kunieda}, \citenamefont
  {Minato},\ and\ \citenamefont {Shibata}}]{CCONE}%
  \BibitemOpen
  \bibfield  {author} {\bibinfo {author} {\bibfnamefont {O.}~\bibnamefont
  {Iwamoto}}, \bibinfo {author} {\bibfnamefont {N.}~\bibnamefont {Iwamoto}},
  \bibinfo {author} {\bibfnamefont {S.}~\bibnamefont {Kunieda}}, \bibinfo
  {author} {\bibfnamefont {F.}~\bibnamefont {Minato}}, \ and\ \bibinfo {author}
  {\bibfnamefont {K.}~\bibnamefont {Shibata}},\ }\href {\doibase
  10.1016/j.nds.2015.12.004} {\bibfield  {journal} {\bibinfo  {journal} {Nucl.
  Data Sheets}\ }\textbf {\bibinfo {volume} {131}},\ \bibinfo {pages} {259}
  (\bibinfo {year} {2016})}\BibitemShut {NoStop}%
\bibitem [{\citenamefont {Koning}\ and\ \citenamefont
  {Delaroche}(2003)}]{KandD2003}%
  \BibitemOpen
  \bibfield  {author} {\bibinfo {author} {\bibfnamefont {A.}~\bibnamefont
  {Koning}}\ and\ \bibinfo {author} {\bibfnamefont {J.}~\bibnamefont
  {Delaroche}},\ }\href@noop {} {\bibfield  {journal} {\bibinfo  {journal}
  {Nucl. Phys.}\ }\textbf {\bibinfo {volume} {A713}} (\bibinfo {year}
  {2003})}\BibitemShut {NoStop}%
\bibitem [{\citenamefont {Avrigeanu}\ and\ \citenamefont
  {Avrigeanu}(2010)}]{Avrigeanu2010}%
  \BibitemOpen
  \bibfield  {author} {\bibinfo {author} {\bibfnamefont {M.}~\bibnamefont
  {Avrigeanu}}\ and\ \bibinfo {author} {\bibfnamefont {V.}~\bibnamefont
  {Avrigeanu}},\ }\href {\doibase 10.1103/PhysRevC.82.014606} {\bibfield
  {journal} {\bibinfo  {journal} {Phys. Rev. C}\ }\textbf {\bibinfo {volume}
  {82}},\ \bibinfo {pages} {014606} (\bibinfo {year} {2010})}\BibitemShut
  {NoStop}%
\bibitem [{\citenamefont {Kopecky}\ and\ \citenamefont {Uhl}(1990)}]{KandU}%
  \BibitemOpen
  \bibfield  {author} {\bibinfo {author} {\bibfnamefont {J.}~\bibnamefont
  {Kopecky}}\ and\ \bibinfo {author} {\bibfnamefont {M.}~\bibnamefont {Uhl}},\
  }\href {\doibase 10.1103/PhysRevC.41.1941} {\bibfield  {journal} {\bibinfo
  {journal} {Phys. Rev. C}\ }\textbf {\bibinfo {volume} {41}},\ \bibinfo
  {pages} {1941} (\bibinfo {year} {1990})}\BibitemShut {NoStop}%
\bibitem [{\citenamefont {Gilbert}\ and\ \citenamefont
  {Cameron}(1965)}]{GilbertCameron}%
  \BibitemOpen
  \bibfield  {author} {\bibinfo {author} {\bibfnamefont {A.}~\bibnamefont
  {Gilbert}}\ and\ \bibinfo {author} {\bibfnamefont {A.~G.~W.}\ \bibnamefont
  {Cameron}},\ }\href {\doibase 10.1139/p65-139} {\bibfield  {journal}
  {\bibinfo  {journal} {Can. J. Phys.}\ }\textbf {\bibinfo {volume} {43}},\
  \bibinfo {pages} {1446} (\bibinfo {year} {1965})}\BibitemShut {NoStop}%
\bibitem [{\citenamefont {Mengoni}\ and\ \citenamefont
  {Nakajima}(1994)}]{MengoniNakajima}%
  \BibitemOpen
  \bibfield  {author} {\bibinfo {author} {\bibfnamefont {A.}~\bibnamefont
  {Mengoni}}\ and\ \bibinfo {author} {\bibfnamefont {Y.}~\bibnamefont
  {Nakajima}},\ }\href {\doibase 10.1080/18811248.1994.9735131} {\bibfield
  {journal} {\bibinfo  {journal} {J. Nucl. Sci. Technol.}\ }\textbf {\bibinfo
  {volume} {31}},\ \bibinfo {pages} {151} (\bibinfo {year} {1994})}\BibitemShut
  {NoStop}%
\bibitem [{\citenamefont {Huang}\ \emph {et~al.}(2017)\citenamefont {Huang},
  \citenamefont {Audi}, \citenamefont {Wang}, \citenamefont {Kondev},
  \citenamefont {Naimi},\ and\ \citenamefont {Xu}}]{AME2016}%
  \BibitemOpen
  \bibfield  {author} {\bibinfo {author} {\bibfnamefont {W.}~\bibnamefont
  {Huang}}, \bibinfo {author} {\bibfnamefont {G.}~\bibnamefont {Audi}},
  \bibinfo {author} {\bibfnamefont {M.}~\bibnamefont {Wang}}, \bibinfo {author}
  {\bibfnamefont {F.~G.}\ \bibnamefont {Kondev}}, \bibinfo {author}
  {\bibfnamefont {S.}~\bibnamefont {Naimi}}, \ and\ \bibinfo {author}
  {\bibfnamefont {X.}~\bibnamefont {Xu}},\ }\href {\doibase
  10.1088/1674-1137/41/3/030002} {\bibfield  {journal} {\bibinfo  {journal}
  {Chin. Phys. C}\ }\textbf {\bibinfo {volume} {41}},\ \bibinfo {pages}
  {030002} (\bibinfo {year} {2017})}\BibitemShut {NoStop}%
\bibitem [{\citenamefont {Wang}\ \emph {et~al.}(2017)\citenamefont {Wang},
  \citenamefont {Audi}, \citenamefont {Kondev}, \citenamefont {Huang},
  \citenamefont {Naimi},\ and\ \citenamefont {Xu}}]{AME2016b}%
  \BibitemOpen
  \bibfield  {author} {\bibinfo {author} {\bibfnamefont {M.}~\bibnamefont
  {Wang}}, \bibinfo {author} {\bibfnamefont {G.}~\bibnamefont {Audi}}, \bibinfo
  {author} {\bibfnamefont {F.~G.}\ \bibnamefont {Kondev}}, \bibinfo {author}
  {\bibfnamefont {W.}~\bibnamefont {Huang}}, \bibinfo {author} {\bibfnamefont
  {S.}~\bibnamefont {Naimi}}, \ and\ \bibinfo {author} {\bibfnamefont
  {X.}~\bibnamefont {Xu}},\ }\href {\doibase 10.1088/1674-1137/41/3/030003}
  {\bibfield  {journal} {\bibinfo  {journal} {Chin. Phys. C}\ }\textbf
  {\bibinfo {volume} {41}},\ \bibinfo {pages} {030003} (\bibinfo {year}
  {2017})}\BibitemShut {NoStop}%
\bibitem [{\citenamefont {M\"oller}\ \emph {et~al.}(2012)\citenamefont
  {M\"oller}, \citenamefont {Myers}, \citenamefont {Sagawa},\ and\
  \citenamefont {Yoshida}}]{FRDM2012}%
  \BibitemOpen
  \bibfield  {author} {\bibinfo {author} {\bibfnamefont {P.}~\bibnamefont
  {M\"oller}}, \bibinfo {author} {\bibfnamefont {W.~D.}\ \bibnamefont {Myers}},
  \bibinfo {author} {\bibfnamefont {H.}~\bibnamefont {Sagawa}}, \ and\ \bibinfo
  {author} {\bibfnamefont {S.}~\bibnamefont {Yoshida}},\ }\href {\doibase
  10.1103/PhysRevLett.108.052501} {\bibfield  {journal} {\bibinfo  {journal}
  {Phys. Rev. Lett.}\ }\textbf {\bibinfo {volume} {108}},\ \bibinfo {pages}
  {052501} (\bibinfo {year} {2012})}\BibitemShut {NoStop}%
\bibitem [{\citenamefont {Lifshitz}\ and\ \citenamefont
  {Singer}(1980)}]{PhysRevC.22.2135}%
  \BibitemOpen
  \bibfield  {author} {\bibinfo {author} {\bibfnamefont {M.}~\bibnamefont
  {Lifshitz}}\ and\ \bibinfo {author} {\bibfnamefont {P.}~\bibnamefont
  {Singer}},\ }\href {\doibase 10.1103/PhysRevC.22.2135} {\bibfield  {journal}
  {\bibinfo  {journal} {Phys. Rev. C}\ }\textbf {\bibinfo {volume} {22}},\
  \bibinfo {pages} {2135} (\bibinfo {year} {1980})}\BibitemShut {NoStop}%
\bibitem [{\citenamefont {Suzuki}\ \emph {et~al.}(1987)\citenamefont {Suzuki},
  \citenamefont {Measday},\ and\ \citenamefont {Roalsvig}}]{Suzuki1987}%
  \BibitemOpen
  \bibfield  {author} {\bibinfo {author} {\bibfnamefont {T.}~\bibnamefont
  {Suzuki}}, \bibinfo {author} {\bibfnamefont {D.~F.}\ \bibnamefont {Measday}},
  \ and\ \bibinfo {author} {\bibfnamefont {J.~P.}\ \bibnamefont {Roalsvig}},\
  }\href {\doibase 10.1103/PhysRevC.35.2212} {\bibfield  {journal} {\bibinfo
  {journal} {Phys. Rev. C}\ }\textbf {\bibinfo {volume} {35}},\ \bibinfo
  {pages} {2212} (\bibinfo {year} {1987})}\BibitemShut {NoStop}%
\bibitem [{\citenamefont {Tanabashi}\ \emph {et~al.}(2018)\citenamefont
  {Tanabashi}, \citenamefont {Hagiwara}, \citenamefont {Hikasa}, \citenamefont
  {Nakamura}, \citenamefont {Sumino}, \citenamefont {Takahashi}, \citenamefont
  {Tanaka}, \citenamefont {Agashe}, \citenamefont {Aielli}, \citenamefont
  {Amsler}, \citenamefont {Antonelli}, \citenamefont {Asner}, \citenamefont
  {Baer}, \citenamefont {Banerjee}, \citenamefont {Barnett}, \citenamefont
  {Basaglia}, \citenamefont {Bauer}, \citenamefont {Beatty}, \citenamefont
  {Belousov}, \citenamefont {Beringer}, \citenamefont {Bethke}, \citenamefont
  {Bettini}, \citenamefont {Bichsel}, \citenamefont {Biebel}, \citenamefont
  {Black}, \citenamefont {Blucher}, \citenamefont {Buchmuller}, \citenamefont
  {Burkert}, \citenamefont {Bychkov}, \citenamefont {Cahn}, \citenamefont
  {Carena}, \citenamefont {Ceccucci}, \citenamefont {Cerri}, \citenamefont
  {Chakraborty}, \citenamefont {Chen}, \citenamefont {Chivukula}, \citenamefont
  {Cowan}, \citenamefont {Dahl}, \citenamefont {D'Ambrosio}, \citenamefont
  {Damour}, \citenamefont {de~Florian}, \citenamefont {de~Gouv\^ea},
  \citenamefont {DeGrand}, \citenamefont {de~Jong}, \citenamefont {Dissertori},
  \citenamefont {Dobrescu}, \citenamefont {D'Onofrio}, \citenamefont {Doser},
  \citenamefont {Drees}, \citenamefont {Dreiner}, \citenamefont {Dwyer},
  \citenamefont {Eerola}, \citenamefont {Eidelman}, \citenamefont {Ellis},
  \citenamefont {Erler}, \citenamefont {Ezhela}, \citenamefont {Fetscher},
  \citenamefont {Fields}, \citenamefont {Firestone}, \citenamefont {Foster},
  \citenamefont {Freitas}, \citenamefont {Gallagher}, \citenamefont {Garren},
  \citenamefont {Gerber}, \citenamefont {Gerbier}, \citenamefont {Gershon},
  \citenamefont {Gershtein}, \citenamefont {Gherghetta}, \citenamefont
  {Godizov}, \citenamefont {Goodman}, \citenamefont {Grab}, \citenamefont
  {Gritsan}, \citenamefont {Grojean}, \citenamefont {Groom}, \citenamefont
  {Gr\"unewald}, \citenamefont {Gurtu}, \citenamefont {Gutsche}, \citenamefont
  {Haber}, \citenamefont {Hanhart}, \citenamefont {Hashimoto}, \citenamefont
  {Hayato}, \citenamefont {Hayes}, \citenamefont {Hebecker}, \citenamefont
  {Heinemeyer}, \citenamefont {Heltsley}, \citenamefont {Hern\'andez-Rey},
  \citenamefont {Hisano}, \citenamefont {H\"ocker}, \citenamefont {Holder},
  \citenamefont {Holtkamp}, \citenamefont {Hyodo}, \citenamefont {Irwin},
  \citenamefont {Johnson}, \citenamefont {Kado}, \citenamefont {Karliner},
  \citenamefont {Katz}, \citenamefont {Klein}, \citenamefont {Klempt},
  \citenamefont {Kowalewski}, \citenamefont {Krauss}, \citenamefont {Kreps},
  \citenamefont {Krusche}, \citenamefont {Kuyanov}, \citenamefont {Kwon},
  \citenamefont {Lahav}, \citenamefont {Laiho}, \citenamefont {Lesgourgues},
  \citenamefont {Liddle}, \citenamefont {Ligeti}, \citenamefont {Lin},
  \citenamefont {Lippmann}, \citenamefont {Liss}, \citenamefont {Littenberg},
  \citenamefont {Lugovsky}, \citenamefont {Lugovsky}, \citenamefont {Lusiani},
  \citenamefont {Makida}, \citenamefont {Maltoni}, \citenamefont {Mannel},
  \citenamefont {Manohar}, \citenamefont {Marciano}, \citenamefont {Martin},
  \citenamefont {Masoni}, \citenamefont {Matthews}, \citenamefont
  {Mei\ss{}ner}, \citenamefont {Milstead}, \citenamefont {Mitchell},
  \citenamefont {M\"onig}, \citenamefont {Molaro}, \citenamefont {Moortgat},
  \citenamefont {Moskovic}, \citenamefont {Murayama}, \citenamefont {Narain},
  \citenamefont {Nason}, \citenamefont {Navas}, \citenamefont {Neubert},
  \citenamefont {Nevski}, \citenamefont {Nir}, \citenamefont {Olive},
  \citenamefont {Pagan~Griso}, \citenamefont {Parsons}, \citenamefont
  {Patrignani}, \citenamefont {Peacock}, \citenamefont {Pennington},
  \citenamefont {Petcov}, \citenamefont {Petrov}, \citenamefont {Pianori},
  \citenamefont {Piepke}, \citenamefont {Pomarol}, \citenamefont {Quadt},
  \citenamefont {Rademacker}, \citenamefont {Raffelt}, \citenamefont
  {Ratcliff}, \citenamefont {Richardson}, \citenamefont {Ringwald},
  \citenamefont {Roesler}, \citenamefont {Rolli}, \citenamefont {Romaniouk},
  \citenamefont {Rosenberg}, \citenamefont {Rosner}, \citenamefont {Rybka},
  \citenamefont {Ryutin}, \citenamefont {Sachrajda}, \citenamefont {Sakai},
  \citenamefont {Salam}, \citenamefont {Sarkar}, \citenamefont {Sauli},
  \citenamefont {Schneider}, \citenamefont {Scholberg}, \citenamefont
  {Schwartz}, \citenamefont {Scott}, \citenamefont {Sharma}, \citenamefont
  {Sharpe}, \citenamefont {Shutt}, \citenamefont {Silari}, \citenamefont
  {Sj\"ostrand}, \citenamefont {Skands}, \citenamefont {Skwarnicki},
  \citenamefont {Smith}, \citenamefont {Smoot}, \citenamefont {Spanier},
  \citenamefont {Spieler}, \citenamefont {Spiering}, \citenamefont {Stahl},
  \citenamefont {Stone}, \citenamefont {Sumiyoshi}, \citenamefont {Syphers},
  \citenamefont {Terashi}, \citenamefont {Terning}, \citenamefont {Thoma},
  \citenamefont {Thorne}, \citenamefont {Tiator}, \citenamefont {Titov},
  \citenamefont {Tkachenko}, \citenamefont {T\"ornqvist}, \citenamefont
  {Tovey}, \citenamefont {Valencia}, \citenamefont {Van~de Water},
  \citenamefont {Varelas}, \citenamefont {Venanzoni}, \citenamefont {Verde},
  \citenamefont {Vincter}, \citenamefont {Vogel}, \citenamefont {Vogt},
  \citenamefont {Wakely}, \citenamefont {Walkowiak}, \citenamefont {Walter},
  \citenamefont {Wands}, \citenamefont {Ward}, \citenamefont {Wascko},
  \citenamefont {Weiglein}, \citenamefont {Weinberg}, \citenamefont {Weinberg},
  \citenamefont {White}, \citenamefont {Wiencke}, \citenamefont {Willocq},
  \citenamefont {Wohl}, \citenamefont {Womersley}, \citenamefont {Woody},
  \citenamefont {Workman}, \citenamefont {Yao}, \citenamefont {Zeller},
  \citenamefont {Zenin}, \citenamefont {Zhu}, \citenamefont {Zhu},
  \citenamefont {Zimmermann}, \citenamefont {Zyla}, \citenamefont {Anderson},
  \citenamefont {Fuller}, \citenamefont {Lugovsky},\ and\ \citenamefont
  {Schaffner}}]{Tanabashi2018Phys.Rev.D98_030001}%
  \BibitemOpen
  \bibfield  {author} {\bibinfo {author} {\bibfnamefont {M.}~\bibnamefont
  {Tanabashi}}, \bibinfo {author} {\bibfnamefont {K.}~\bibnamefont {Hagiwara}},
  \bibinfo {author} {\bibfnamefont {K.}~\bibnamefont {Hikasa}}, \bibinfo
  {author} {\bibfnamefont {K.}~\bibnamefont {Nakamura}}, \bibinfo {author}
  {\bibfnamefont {Y.}~\bibnamefont {Sumino}}, \bibinfo {author} {\bibfnamefont
  {F.}~\bibnamefont {Takahashi}}, \bibinfo {author} {\bibfnamefont
  {J.}~\bibnamefont {Tanaka}}, \bibinfo {author} {\bibfnamefont
  {K.}~\bibnamefont {Agashe}}, \bibinfo {author} {\bibfnamefont
  {G.}~\bibnamefont {Aielli}}, \bibinfo {author} {\bibfnamefont
  {C.}~\bibnamefont {Amsler}}, \bibinfo {author} {\bibfnamefont
  {M.}~\bibnamefont {Antonelli}}, \bibinfo {author} {\bibfnamefont {D.~M.}\
  \bibnamefont {Asner}}, \bibinfo {author} {\bibfnamefont {H.}~\bibnamefont
  {Baer}}, \bibinfo {author} {\bibfnamefont {S.}~\bibnamefont {Banerjee}},
  \bibinfo {author} {\bibfnamefont {R.~M.}\ \bibnamefont {Barnett}}, \bibinfo
  {author} {\bibfnamefont {T.}~\bibnamefont {Basaglia}}, \bibinfo {author}
  {\bibfnamefont {C.~W.}\ \bibnamefont {Bauer}}, \bibinfo {author}
  {\bibfnamefont {J.~J.}\ \bibnamefont {Beatty}}, \bibinfo {author}
  {\bibfnamefont {V.~I.}\ \bibnamefont {Belousov}}, \bibinfo {author}
  {\bibfnamefont {J.}~\bibnamefont {Beringer}}, \bibinfo {author}
  {\bibfnamefont {S.}~\bibnamefont {Bethke}}, \bibinfo {author} {\bibfnamefont
  {A.}~\bibnamefont {Bettini}}, \bibinfo {author} {\bibfnamefont
  {H.}~\bibnamefont {Bichsel}}, \bibinfo {author} {\bibfnamefont
  {O.}~\bibnamefont {Biebel}}, \bibinfo {author} {\bibfnamefont {K.~M.}\
  \bibnamefont {Black}}, \bibinfo {author} {\bibfnamefont {E.}~\bibnamefont
  {Blucher}}, \bibinfo {author} {\bibfnamefont {O.}~\bibnamefont {Buchmuller}},
  \bibinfo {author} {\bibfnamefont {V.}~\bibnamefont {Burkert}}, \bibinfo
  {author} {\bibfnamefont {M.~A.}\ \bibnamefont {Bychkov}}, \bibinfo {author}
  {\bibfnamefont {R.~N.}\ \bibnamefont {Cahn}}, \bibinfo {author}
  {\bibfnamefont {M.}~\bibnamefont {Carena}}, \bibinfo {author} {\bibfnamefont
  {A.}~\bibnamefont {Ceccucci}}, \bibinfo {author} {\bibfnamefont
  {A.}~\bibnamefont {Cerri}}, \bibinfo {author} {\bibfnamefont
  {D.}~\bibnamefont {Chakraborty}}, \bibinfo {author} {\bibfnamefont {M.-C.}\
  \bibnamefont {Chen}}, \bibinfo {author} {\bibfnamefont {R.~S.}\ \bibnamefont
  {Chivukula}}, \bibinfo {author} {\bibfnamefont {G.}~\bibnamefont {Cowan}},
  \bibinfo {author} {\bibfnamefont {O.}~\bibnamefont {Dahl}}, \bibinfo {author}
  {\bibfnamefont {G.}~\bibnamefont {D'Ambrosio}}, \bibinfo {author}
  {\bibfnamefont {T.}~\bibnamefont {Damour}}, \bibinfo {author} {\bibfnamefont
  {D.}~\bibnamefont {de~Florian}}, \bibinfo {author} {\bibfnamefont
  {A.}~\bibnamefont {de~Gouv\^ea}}, \bibinfo {author} {\bibfnamefont
  {T.}~\bibnamefont {DeGrand}}, \bibinfo {author} {\bibfnamefont
  {P.}~\bibnamefont {de~Jong}}, \bibinfo {author} {\bibfnamefont
  {G.}~\bibnamefont {Dissertori}}, \bibinfo {author} {\bibfnamefont {B.~A.}\
  \bibnamefont {Dobrescu}}, \bibinfo {author} {\bibfnamefont {M.}~\bibnamefont
  {D'Onofrio}}, \bibinfo {author} {\bibfnamefont {M.}~\bibnamefont {Doser}},
  \bibinfo {author} {\bibfnamefont {M.}~\bibnamefont {Drees}}, \bibinfo
  {author} {\bibfnamefont {H.~K.}\ \bibnamefont {Dreiner}}, \bibinfo {author}
  {\bibfnamefont {D.~A.}\ \bibnamefont {Dwyer}}, \bibinfo {author}
  {\bibfnamefont {P.}~\bibnamefont {Eerola}}, \bibinfo {author} {\bibfnamefont
  {S.}~\bibnamefont {Eidelman}}, \bibinfo {author} {\bibfnamefont
  {J.}~\bibnamefont {Ellis}}, \bibinfo {author} {\bibfnamefont
  {J.}~\bibnamefont {Erler}}, \bibinfo {author} {\bibfnamefont {V.~V.}\
  \bibnamefont {Ezhela}}, \bibinfo {author} {\bibfnamefont {W.}~\bibnamefont
  {Fetscher}}, \bibinfo {author} {\bibfnamefont {B.~D.}\ \bibnamefont
  {Fields}}, \bibinfo {author} {\bibfnamefont {R.}~\bibnamefont {Firestone}},
  \bibinfo {author} {\bibfnamefont {B.}~\bibnamefont {Foster}}, \bibinfo
  {author} {\bibfnamefont {A.}~\bibnamefont {Freitas}}, \bibinfo {author}
  {\bibfnamefont {H.}~\bibnamefont {Gallagher}}, \bibinfo {author}
  {\bibfnamefont {L.}~\bibnamefont {Garren}}, \bibinfo {author} {\bibfnamefont
  {H.-J.}\ \bibnamefont {Gerber}}, \bibinfo {author} {\bibfnamefont
  {G.}~\bibnamefont {Gerbier}}, \bibinfo {author} {\bibfnamefont
  {T.}~\bibnamefont {Gershon}}, \bibinfo {author} {\bibfnamefont
  {Y.}~\bibnamefont {Gershtein}}, \bibinfo {author} {\bibfnamefont
  {T.}~\bibnamefont {Gherghetta}}, \bibinfo {author} {\bibfnamefont {A.~A.}\
  \bibnamefont {Godizov}}, \bibinfo {author} {\bibfnamefont {M.}~\bibnamefont
  {Goodman}}, \bibinfo {author} {\bibfnamefont {C.}~\bibnamefont {Grab}},
  \bibinfo {author} {\bibfnamefont {A.~V.}\ \bibnamefont {Gritsan}}, \bibinfo
  {author} {\bibfnamefont {C.}~\bibnamefont {Grojean}}, \bibinfo {author}
  {\bibfnamefont {D.~E.}\ \bibnamefont {Groom}}, \bibinfo {author}
  {\bibfnamefont {M.}~\bibnamefont {Gr\"unewald}}, \bibinfo {author}
  {\bibfnamefont {A.}~\bibnamefont {Gurtu}}, \bibinfo {author} {\bibfnamefont
  {T.}~\bibnamefont {Gutsche}}, \bibinfo {author} {\bibfnamefont {H.~E.}\
  \bibnamefont {Haber}}, \bibinfo {author} {\bibfnamefont {C.}~\bibnamefont
  {Hanhart}}, \bibinfo {author} {\bibfnamefont {S.}~\bibnamefont {Hashimoto}},
  \bibinfo {author} {\bibfnamefont {Y.}~\bibnamefont {Hayato}}, \bibinfo
  {author} {\bibfnamefont {K.~G.}\ \bibnamefont {Hayes}}, \bibinfo {author}
  {\bibfnamefont {A.}~\bibnamefont {Hebecker}}, \bibinfo {author}
  {\bibfnamefont {S.}~\bibnamefont {Heinemeyer}}, \bibinfo {author}
  {\bibfnamefont {B.}~\bibnamefont {Heltsley}}, \bibinfo {author}
  {\bibfnamefont {J.~J.}\ \bibnamefont {Hern\'andez-Rey}}, \bibinfo {author}
  {\bibfnamefont {J.}~\bibnamefont {Hisano}}, \bibinfo {author} {\bibfnamefont
  {A.}~\bibnamefont {H\"ocker}}, \bibinfo {author} {\bibfnamefont
  {J.}~\bibnamefont {Holder}}, \bibinfo {author} {\bibfnamefont
  {A.}~\bibnamefont {Holtkamp}}, \bibinfo {author} {\bibfnamefont
  {T.}~\bibnamefont {Hyodo}}, \bibinfo {author} {\bibfnamefont {K.~D.}\
  \bibnamefont {Irwin}}, \bibinfo {author} {\bibfnamefont {K.~F.}\ \bibnamefont
  {Johnson}}, \bibinfo {author} {\bibfnamefont {M.}~\bibnamefont {Kado}},
  \bibinfo {author} {\bibfnamefont {M.}~\bibnamefont {Karliner}}, \bibinfo
  {author} {\bibfnamefont {U.~F.}\ \bibnamefont {Katz}}, \bibinfo {author}
  {\bibfnamefont {S.~R.}\ \bibnamefont {Klein}}, \bibinfo {author}
  {\bibfnamefont {E.}~\bibnamefont {Klempt}}, \bibinfo {author} {\bibfnamefont
  {R.~V.}\ \bibnamefont {Kowalewski}}, \bibinfo {author} {\bibfnamefont
  {F.}~\bibnamefont {Krauss}}, \bibinfo {author} {\bibfnamefont
  {M.}~\bibnamefont {Kreps}}, \bibinfo {author} {\bibfnamefont
  {B.}~\bibnamefont {Krusche}}, \bibinfo {author} {\bibfnamefont {Y.~V.}\
  \bibnamefont {Kuyanov}}, \bibinfo {author} {\bibfnamefont {Y.}~\bibnamefont
  {Kwon}}, \bibinfo {author} {\bibfnamefont {O.}~\bibnamefont {Lahav}},
  \bibinfo {author} {\bibfnamefont {J.}~\bibnamefont {Laiho}}, \bibinfo
  {author} {\bibfnamefont {J.}~\bibnamefont {Lesgourgues}}, \bibinfo {author}
  {\bibfnamefont {A.}~\bibnamefont {Liddle}}, \bibinfo {author} {\bibfnamefont
  {Z.}~\bibnamefont {Ligeti}}, \bibinfo {author} {\bibfnamefont {C.-J.}\
  \bibnamefont {Lin}}, \bibinfo {author} {\bibfnamefont {C.}~\bibnamefont
  {Lippmann}}, \bibinfo {author} {\bibfnamefont {T.~M.}\ \bibnamefont {Liss}},
  \bibinfo {author} {\bibfnamefont {L.}~\bibnamefont {Littenberg}}, \bibinfo
  {author} {\bibfnamefont {K.~S.}\ \bibnamefont {Lugovsky}}, \bibinfo {author}
  {\bibfnamefont {S.~B.}\ \bibnamefont {Lugovsky}}, \bibinfo {author}
  {\bibfnamefont {A.}~\bibnamefont {Lusiani}}, \bibinfo {author} {\bibfnamefont
  {Y.}~\bibnamefont {Makida}}, \bibinfo {author} {\bibfnamefont
  {F.}~\bibnamefont {Maltoni}}, \bibinfo {author} {\bibfnamefont
  {T.}~\bibnamefont {Mannel}}, \bibinfo {author} {\bibfnamefont {A.~V.}\
  \bibnamefont {Manohar}}, \bibinfo {author} {\bibfnamefont {W.~J.}\
  \bibnamefont {Marciano}}, \bibinfo {author} {\bibfnamefont {A.~D.}\
  \bibnamefont {Martin}}, \bibinfo {author} {\bibfnamefont {A.}~\bibnamefont
  {Masoni}}, \bibinfo {author} {\bibfnamefont {J.}~\bibnamefont {Matthews}},
  \bibinfo {author} {\bibfnamefont {U.-G.}\ \bibnamefont {Mei\ss{}ner}},
  \bibinfo {author} {\bibfnamefont {D.}~\bibnamefont {Milstead}}, \bibinfo
  {author} {\bibfnamefont {R.~E.}\ \bibnamefont {Mitchell}}, \bibinfo {author}
  {\bibfnamefont {K.}~\bibnamefont {M\"onig}}, \bibinfo {author} {\bibfnamefont
  {P.}~\bibnamefont {Molaro}}, \bibinfo {author} {\bibfnamefont
  {F.}~\bibnamefont {Moortgat}}, \bibinfo {author} {\bibfnamefont
  {M.}~\bibnamefont {Moskovic}}, \bibinfo {author} {\bibfnamefont
  {H.}~\bibnamefont {Murayama}}, \bibinfo {author} {\bibfnamefont
  {M.}~\bibnamefont {Narain}}, \bibinfo {author} {\bibfnamefont
  {P.}~\bibnamefont {Nason}}, \bibinfo {author} {\bibfnamefont
  {S.}~\bibnamefont {Navas}}, \bibinfo {author} {\bibfnamefont
  {M.}~\bibnamefont {Neubert}}, \bibinfo {author} {\bibfnamefont
  {P.}~\bibnamefont {Nevski}}, \bibinfo {author} {\bibfnamefont
  {Y.}~\bibnamefont {Nir}}, \bibinfo {author} {\bibfnamefont {K.~A.}\
  \bibnamefont {Olive}}, \bibinfo {author} {\bibfnamefont {S.}~\bibnamefont
  {Pagan~Griso}}, \bibinfo {author} {\bibfnamefont {J.}~\bibnamefont
  {Parsons}}, \bibinfo {author} {\bibfnamefont {C.}~\bibnamefont {Patrignani}},
  \bibinfo {author} {\bibfnamefont {J.~A.}\ \bibnamefont {Peacock}}, \bibinfo
  {author} {\bibfnamefont {M.}~\bibnamefont {Pennington}}, \bibinfo {author}
  {\bibfnamefont {S.~T.}\ \bibnamefont {Petcov}}, \bibinfo {author}
  {\bibfnamefont {V.~A.}\ \bibnamefont {Petrov}}, \bibinfo {author}
  {\bibfnamefont {E.}~\bibnamefont {Pianori}}, \bibinfo {author} {\bibfnamefont
  {A.}~\bibnamefont {Piepke}}, \bibinfo {author} {\bibfnamefont
  {A.}~\bibnamefont {Pomarol}}, \bibinfo {author} {\bibfnamefont
  {A.}~\bibnamefont {Quadt}}, \bibinfo {author} {\bibfnamefont
  {J.}~\bibnamefont {Rademacker}}, \bibinfo {author} {\bibfnamefont
  {G.}~\bibnamefont {Raffelt}}, \bibinfo {author} {\bibfnamefont {B.~N.}\
  \bibnamefont {Ratcliff}}, \bibinfo {author} {\bibfnamefont {P.}~\bibnamefont
  {Richardson}}, \bibinfo {author} {\bibfnamefont {A.}~\bibnamefont
  {Ringwald}}, \bibinfo {author} {\bibfnamefont {S.}~\bibnamefont {Roesler}},
  \bibinfo {author} {\bibfnamefont {S.}~\bibnamefont {Rolli}}, \bibinfo
  {author} {\bibfnamefont {A.}~\bibnamefont {Romaniouk}}, \bibinfo {author}
  {\bibfnamefont {L.~J.}\ \bibnamefont {Rosenberg}}, \bibinfo {author}
  {\bibfnamefont {J.~L.}\ \bibnamefont {Rosner}}, \bibinfo {author}
  {\bibfnamefont {G.}~\bibnamefont {Rybka}}, \bibinfo {author} {\bibfnamefont
  {R.~A.}\ \bibnamefont {Ryutin}}, \bibinfo {author} {\bibfnamefont {C.~T.}\
  \bibnamefont {Sachrajda}}, \bibinfo {author} {\bibfnamefont {Y.}~\bibnamefont
  {Sakai}}, \bibinfo {author} {\bibfnamefont {G.~P.}\ \bibnamefont {Salam}},
  \bibinfo {author} {\bibfnamefont {S.}~\bibnamefont {Sarkar}}, \bibinfo
  {author} {\bibfnamefont {F.}~\bibnamefont {Sauli}}, \bibinfo {author}
  {\bibfnamefont {O.}~\bibnamefont {Schneider}}, \bibinfo {author}
  {\bibfnamefont {K.}~\bibnamefont {Scholberg}}, \bibinfo {author}
  {\bibfnamefont {A.~J.}\ \bibnamefont {Schwartz}}, \bibinfo {author}
  {\bibfnamefont {D.}~\bibnamefont {Scott}}, \bibinfo {author} {\bibfnamefont
  {V.}~\bibnamefont {Sharma}}, \bibinfo {author} {\bibfnamefont {S.~R.}\
  \bibnamefont {Sharpe}}, \bibinfo {author} {\bibfnamefont {T.}~\bibnamefont
  {Shutt}}, \bibinfo {author} {\bibfnamefont {M.}~\bibnamefont {Silari}},
  \bibinfo {author} {\bibfnamefont {T.}~\bibnamefont {Sj\"ostrand}}, \bibinfo
  {author} {\bibfnamefont {P.}~\bibnamefont {Skands}}, \bibinfo {author}
  {\bibfnamefont {T.}~\bibnamefont {Skwarnicki}}, \bibinfo {author}
  {\bibfnamefont {J.~G.}\ \bibnamefont {Smith}}, \bibinfo {author}
  {\bibfnamefont {G.~F.}\ \bibnamefont {Smoot}}, \bibinfo {author}
  {\bibfnamefont {S.}~\bibnamefont {Spanier}}, \bibinfo {author} {\bibfnamefont
  {H.}~\bibnamefont {Spieler}}, \bibinfo {author} {\bibfnamefont
  {C.}~\bibnamefont {Spiering}}, \bibinfo {author} {\bibfnamefont
  {A.}~\bibnamefont {Stahl}}, \bibinfo {author} {\bibfnamefont {S.~L.}\
  \bibnamefont {Stone}}, \bibinfo {author} {\bibfnamefont {T.}~\bibnamefont
  {Sumiyoshi}}, \bibinfo {author} {\bibfnamefont {M.~J.}\ \bibnamefont
  {Syphers}}, \bibinfo {author} {\bibfnamefont {K.}~\bibnamefont {Terashi}},
  \bibinfo {author} {\bibfnamefont {J.}~\bibnamefont {Terning}}, \bibinfo
  {author} {\bibfnamefont {U.}~\bibnamefont {Thoma}}, \bibinfo {author}
  {\bibfnamefont {R.~S.}\ \bibnamefont {Thorne}}, \bibinfo {author}
  {\bibfnamefont {L.}~\bibnamefont {Tiator}}, \bibinfo {author} {\bibfnamefont
  {M.}~\bibnamefont {Titov}}, \bibinfo {author} {\bibfnamefont {N.~P.}\
  \bibnamefont {Tkachenko}}, \bibinfo {author} {\bibfnamefont {N.~A.}\
  \bibnamefont {T\"ornqvist}}, \bibinfo {author} {\bibfnamefont {D.~R.}\
  \bibnamefont {Tovey}}, \bibinfo {author} {\bibfnamefont {G.}~\bibnamefont
  {Valencia}}, \bibinfo {author} {\bibfnamefont {R.}~\bibnamefont {Van~de
  Water}}, \bibinfo {author} {\bibfnamefont {N.}~\bibnamefont {Varelas}},
  \bibinfo {author} {\bibfnamefont {G.}~\bibnamefont {Venanzoni}}, \bibinfo
  {author} {\bibfnamefont {L.}~\bibnamefont {Verde}}, \bibinfo {author}
  {\bibfnamefont {M.~G.}\ \bibnamefont {Vincter}}, \bibinfo {author}
  {\bibfnamefont {P.}~\bibnamefont {Vogel}}, \bibinfo {author} {\bibfnamefont
  {A.}~\bibnamefont {Vogt}}, \bibinfo {author} {\bibfnamefont {S.~P.}\
  \bibnamefont {Wakely}}, \bibinfo {author} {\bibfnamefont {W.}~\bibnamefont
  {Walkowiak}}, \bibinfo {author} {\bibfnamefont {C.~W.}\ \bibnamefont
  {Walter}}, \bibinfo {author} {\bibfnamefont {D.}~\bibnamefont {Wands}},
  \bibinfo {author} {\bibfnamefont {D.~R.}\ \bibnamefont {Ward}}, \bibinfo
  {author} {\bibfnamefont {M.~O.}\ \bibnamefont {Wascko}}, \bibinfo {author}
  {\bibfnamefont {G.}~\bibnamefont {Weiglein}}, \bibinfo {author}
  {\bibfnamefont {D.~H.}\ \bibnamefont {Weinberg}}, \bibinfo {author}
  {\bibfnamefont {E.~J.}\ \bibnamefont {Weinberg}}, \bibinfo {author}
  {\bibfnamefont {M.}~\bibnamefont {White}}, \bibinfo {author} {\bibfnamefont
  {L.~R.}\ \bibnamefont {Wiencke}}, \bibinfo {author} {\bibfnamefont
  {S.}~\bibnamefont {Willocq}}, \bibinfo {author} {\bibfnamefont {C.~G.}\
  \bibnamefont {Wohl}}, \bibinfo {author} {\bibfnamefont {J.}~\bibnamefont
  {Womersley}}, \bibinfo {author} {\bibfnamefont {C.~L.}\ \bibnamefont
  {Woody}}, \bibinfo {author} {\bibfnamefont {R.~L.}\ \bibnamefont {Workman}},
  \bibinfo {author} {\bibfnamefont {W.-M.}\ \bibnamefont {Yao}}, \bibinfo
  {author} {\bibfnamefont {G.~P.}\ \bibnamefont {Zeller}}, \bibinfo {author}
  {\bibfnamefont {O.~V.}\ \bibnamefont {Zenin}}, \bibinfo {author}
  {\bibfnamefont {R.-Y.}\ \bibnamefont {Zhu}}, \bibinfo {author} {\bibfnamefont
  {S.-L.}\ \bibnamefont {Zhu}}, \bibinfo {author} {\bibfnamefont
  {F.}~\bibnamefont {Zimmermann}}, \bibinfo {author} {\bibfnamefont {P.~A.}\
  \bibnamefont {Zyla}}, \bibinfo {author} {\bibfnamefont {J.}~\bibnamefont
  {Anderson}}, \bibinfo {author} {\bibfnamefont {L.}~\bibnamefont {Fuller}},
  \bibinfo {author} {\bibfnamefont {V.~S.}\ \bibnamefont {Lugovsky}}, \ and\
  \bibinfo {author} {\bibfnamefont {P.}~\bibnamefont {Schaffner}} (\bibinfo
  {collaboration} {Particle Data Group}),\ }\href {\doibase
  10.1103/PhysRevD.98.030001} {\bibfield  {journal} {\bibinfo  {journal} {Phys.
  Rev. D}\ }\textbf {\bibinfo {volume} {98}},\ \bibinfo {pages} {030001}
  (\bibinfo {year} {2018})}\BibitemShut {NoStop}%
\bibitem [{\citenamefont {Miyake}\ \emph {et~al.}(2009)\citenamefont {Miyake},
  \citenamefont {Nishiyama}, \citenamefont {Kawamura}, \citenamefont
  {Strasser}, \citenamefont {Makimura}, \citenamefont {Koda}, \citenamefont
  {Shimomura}, \citenamefont {Fujimori}, \citenamefont {Nakahara},
  \citenamefont {Kadono}, \citenamefont {Kato}, \citenamefont {Takeshita},
  \citenamefont {Higemoto}, \citenamefont {Ishida}, \citenamefont {Matsuzaki},
  \citenamefont {Matsuda},\ and\ \citenamefont {Nagamine}}]{J-parc}%
  \BibitemOpen
  \bibfield  {author} {\bibinfo {author} {\bibfnamefont {Y.}~\bibnamefont
  {Miyake}}, \bibinfo {author} {\bibfnamefont {K.}~\bibnamefont {Nishiyama}},
  \bibinfo {author} {\bibfnamefont {N.}~\bibnamefont {Kawamura}}, \bibinfo
  {author} {\bibfnamefont {P.}~\bibnamefont {Strasser}}, \bibinfo {author}
  {\bibfnamefont {S.}~\bibnamefont {Makimura}}, \bibinfo {author}
  {\bibfnamefont {A.}~\bibnamefont {Koda}}, \bibinfo {author} {\bibfnamefont
  {K.}~\bibnamefont {Shimomura}}, \bibinfo {author} {\bibfnamefont
  {H.}~\bibnamefont {Fujimori}}, \bibinfo {author} {\bibfnamefont
  {K.}~\bibnamefont {Nakahara}}, \bibinfo {author} {\bibfnamefont
  {R.}~\bibnamefont {Kadono}}, \bibinfo {author} {\bibfnamefont
  {M.}~\bibnamefont {Kato}}, \bibinfo {author} {\bibfnamefont {S.}~\bibnamefont
  {Takeshita}}, \bibinfo {author} {\bibfnamefont {W.}~\bibnamefont {Higemoto}},
  \bibinfo {author} {\bibfnamefont {K.}~\bibnamefont {Ishida}}, \bibinfo
  {author} {\bibfnamefont {T.}~\bibnamefont {Matsuzaki}}, \bibinfo {author}
  {\bibfnamefont {Y.}~\bibnamefont {Matsuda}}, \ and\ \bibinfo {author}
  {\bibfnamefont {K.}~\bibnamefont {Nagamine}},\ }\href {\doibase
  10.1016/j.nima.2008.11.016} {\bibfield  {journal} {\bibinfo  {journal} {Nucl.
  Instrum. Methods Phys. Res., Sect. A}\ }\textbf {\bibinfo {volume} {600}},\
  \bibinfo {pages} {22} (\bibinfo {year} {2009})}\BibitemShut {NoStop}%
\bibitem [{\citenamefont {Mori}\ \emph {et~al.}(2018)\citenamefont {Mori},
  \citenamefont {Taniguchi}, \citenamefont {Kuriyama}, \citenamefont {Uesugi},
  \citenamefont {Ishi}, \citenamefont {Muto}, \citenamefont {Ono},
  \citenamefont {Okita}, \citenamefont {Sato}, \citenamefont {Kinsho},
  \citenamefont {Miyake}, \citenamefont {Yoshimoto},\ and\ \citenamefont
  {Okabe}}]{Merit}%
  \BibitemOpen
  \bibfield  {author} {\bibinfo {author} {\bibfnamefont {Y.}~\bibnamefont
  {Mori}}, \bibinfo {author} {\bibfnamefont {A.}~\bibnamefont {Taniguchi}},
  \bibinfo {author} {\bibfnamefont {Y.}~\bibnamefont {Kuriyama}}, \bibinfo
  {author} {\bibfnamefont {T.}~\bibnamefont {Uesugi}}, \bibinfo {author}
  {\bibfnamefont {Y.}~\bibnamefont {Ishi}}, \bibinfo {author} {\bibfnamefont
  {M.}~\bibnamefont {Muto}}, \bibinfo {author} {\bibfnamefont {Y.}~\bibnamefont
  {Ono}}, \bibinfo {author} {\bibfnamefont {H.}~\bibnamefont {Okita}}, \bibinfo
  {author} {\bibfnamefont {A.}~\bibnamefont {Sato}}, \bibinfo {author}
  {\bibfnamefont {M.}~\bibnamefont {Kinsho}}, \bibinfo {author} {\bibfnamefont
  {Y.}~\bibnamefont {Miyake}}, \bibinfo {author} {\bibfnamefont
  {M.}~\bibnamefont {Yoshimoto}}, \ and\ \bibinfo {author} {\bibfnamefont
  {K.}~\bibnamefont {Okabe}},\ }\href {\doibase 10.7566/JPSCP.21.011063}
  {\bibfield  {journal} {\bibinfo  {journal} {JPS Conf. Proc.}\ }\textbf
  {\bibinfo {volume} {21}},\ \bibinfo {pages} {011063} (\bibinfo {year}
  {2018})}\BibitemShut {NoStop}%
\bibitem [{\citenamefont {Abe}\ and\ \citenamefont {Sato}(2016)}]{Abe2016}%
  \BibitemOpen
  \bibfield  {author} {\bibinfo {author} {\bibfnamefont {S.-i.}\ \bibnamefont
  {Abe}}\ and\ \bibinfo {author} {\bibfnamefont {T.}~\bibnamefont {Sato}},\
  }\href {\doibase 10.1051/epjconf/201612204002} {\bibfield  {journal}
  {\bibinfo  {journal} {EPJ Web Conf.}\ }\textbf {\bibinfo {volume} {122}},\
  \bibinfo {pages} {04002} (\bibinfo {year} {2016})}\BibitemShut {NoStop}%
\bibitem [{\citenamefont {Sato}\ \emph {et~al.}(2018)\citenamefont {Sato},
  \citenamefont {Iwamoto}, \citenamefont {Hashimoto}, \citenamefont {Ogawa},
  \citenamefont {Furuta}, \citenamefont {ichiro Abe}, \citenamefont {Kai},
  \citenamefont {Tsai}, \citenamefont {Matsuda}, \citenamefont {Iwase},
  \citenamefont {Shigyo}, \citenamefont {Sihver},\ and\ \citenamefont
  {Niita}}]{Sato2018}%
  \BibitemOpen
  \bibfield  {author} {\bibinfo {author} {\bibfnamefont {T.}~\bibnamefont
  {Sato}}, \bibinfo {author} {\bibfnamefont {Y.}~\bibnamefont {Iwamoto}},
  \bibinfo {author} {\bibfnamefont {S.}~\bibnamefont {Hashimoto}}, \bibinfo
  {author} {\bibfnamefont {T.}~\bibnamefont {Ogawa}}, \bibinfo {author}
  {\bibfnamefont {T.}~\bibnamefont {Furuta}}, \bibinfo {author} {\bibfnamefont
  {S.}~\bibnamefont {ichiro Abe}}, \bibinfo {author} {\bibfnamefont
  {T.}~\bibnamefont {Kai}}, \bibinfo {author} {\bibfnamefont {P.-E.}\
  \bibnamefont {Tsai}}, \bibinfo {author} {\bibfnamefont {N.}~\bibnamefont
  {Matsuda}}, \bibinfo {author} {\bibfnamefont {H.}~\bibnamefont {Iwase}},
  \bibinfo {author} {\bibfnamefont {N.}~\bibnamefont {Shigyo}}, \bibinfo
  {author} {\bibfnamefont {L.}~\bibnamefont {Sihver}}, \ and\ \bibinfo {author}
  {\bibfnamefont {K.}~\bibnamefont {Niita}},\ }\href {\doibase
  10.1080/00223131.2017.1419890} {\bibfield  {journal} {\bibinfo  {journal} {J.
  Nucl. Sci. Technol.}\ }\textbf {\bibinfo {volume} {55}},\ \bibinfo {pages}
  {684} (\bibinfo {year} {2018})}\BibitemShut {NoStop}%
\bibitem [{\citenamefont
  {Greiner}(1998)}]{Greiner1998ClassicalElectrodynamics_Springer-Verlag}%
  \BibitemOpen
  \bibfield  {author} {\bibinfo {author} {\bibfnamefont {W.}~\bibnamefont
  {Greiner}},\ }\href@noop {} {\emph {\bibinfo {title} {{Classical
  Electrodynamics}}}}\ (\bibinfo  {publisher} {Springer-Verlag},\ \bibinfo
  {address} {New York},\ \bibinfo {year} {1998})\BibitemShut {NoStop}%
\end{thebibliography}%
\end{document}